\documentclass[useAMS,usenatbib,usegraphicx]{mn2e}

\usepackage{graphicx}
\usepackage{times}
\newcounter{subfigure}
\def\lea{\mathrel{<\kern-1.0em\lower0.9ex\hbox{$\sim$}}}
\def\gea{\mathrel{>\kern-1.0em\lower0.9ex\hbox{$\sim$}}}
\def\leq{\mathrel{<\kern-1.0em\lower0.9ex\hbox{$-$}}}
\def\geq{\mathrel{>\kern-1.0em\lower0.9ex\hbox{$-$}}}

\newcommand{\lesssim}{\mathrel{\hbox{\rlap{\hbox{\lower4pt\hbox{$\sim$}}}\hbox{$<$}}}}
\newcommand{\gtrsim}{\mathrel{\hbox{\rlap{\hbox{\lower4pt\hbox{$\sim$}}}\hbox{$>$}}}}

\title[The FUV excess in early-type galaxies]{The spatial distribution and origin of the FUV excess in early-type galaxies}

\author[David Carter et al.]{David Carter$^1$, Sally Pass$^{1,2}$, Joseph Kennedy$^3$, Arna M. Karick$^1$ \& Russell J. Smith$^4$\newauthor
\\$^{1}$Astrophysics Research Institute, Liverpool John Moores 
University, Twelve Quays House, Egerton Wharf, Birkenhead CH41 1LD,
UK.
\\$^2$Department of Physics, University of Liverpool, Liverpool L69 7ZE,
UK.
\\$^3$Calderstones School, Harthill Road, Allerton, Liverpool, L18 3HS, UK. 
\\$^4$Department of Physics, University of Durham, Durham DH1 3LE, UK.
}

\begin{document}

\date{}

\pagerange{\pageref{firstpage}--\pageref{lastpage}} \pubyear{2009}

\maketitle

\label{firstpage}

\begin{abstract}
We present surface photometry of a sample of 52 galaxies from the GALEX and 2MASS data archives, these include
32 normal elliptical galaxies, 10 ellipticals with weak Liner or other nuclear activity, and 10 star forming ellipticals
or early-type spirals. We examine the spatial distribution of the Far Ultra-Violet excess in these galaxies, and its 
correlation with dynamical and stellar population properties of the galaxies. From aperture photometry we find that 
all galaxies except for recent major remnants and galaxies with ongoing star formation show a positive gradient
in the (FUV-NUV) colour determined from the GALEX images. The logarithmic gradient does not correlate
with any stellar population parameter, but it does correlate with the 
central velocity dispersion. The strength of the excess on the other hand, correlates with both [$\alpha$/Fe] and
[Z/H], but more strongly with the former. We derive models of the underlying stellar population from the 2MASS
H-band images, and the residual of the image from this model reveals a map of the centrally concentrated FUV excess.
We examine a possible hypothesis for generating the FUV excess and the radial gradient in its strength, involving
a helium abundance gradient set up early in the formation process of the galaxies. If this hypothesis is correct, the
persistence of the gradients to the present day places a strong limit on the importance of dry mergers in the formation
of ellipticals.

\end{abstract}

\begin{keywords}
galaxies: stellar content; galaxies: elliptical and lenticular, cD; galaxies: evolution
\end{keywords}

\section{Introduction}

It has long been known that the spectral energy distributions of
bright early-type galaxies have a pronounced upturn at wavelengths
shortward of 2100{\AA} (Code \& Welch 1979; Brown et al. 1997;
O'Connell et al. 1992; O'Connell 1999). This  far-UV excess was
unexpected as early-type galaxies were believed
to contain only old and cold stellar populations. From the shape of
the spectral energy distribution in the UV, it appears that the
temperature of the stars which give rise to the excess has a narrow
range around 20,000$\pm$3000K (Brown et al. 1997) which rules out upper main
sequence stars as an origin.  It is believed to be caused instead
by hot subdwarfs, also known as extreme horizontal branch (EHB) stars
(see Yi 2008 for a review).

A feature of the UV upturn is that it is strongest in the centres of bright
elliptical galaxies (O'Connell et al. 1992; 1999). The positive UV colour gradients
have been interpreted as abundance gradients, but it is still unclear which 
abundance or abundances drive these gradients.

Such stars are found  in globular clusters, and are generally considered
to be helium-core burning stars with extremely thin hydrogen envelopes
($<0.02M_\odot$). There are a number of competing models for their
origin. Han et al.\ (2002, 2003, 2007) and Han (2008), propose a binary model,
in which a star loses all of its envelope near the tip
of the red-giant branch (RGB) by mass
transfer to a companion star or ejection in a common-envelope phase,
or where two helium white dwarfs merge with a combined mass
larger than $\sim 0.35\,M_{\odot}$. In each case, the remnant star 
ignites helium and becomes a hot subdwarf. 

A number of single-star mechanisms have been proposed to explain the
EHB populations of ellipticals. Park \& Lee
(1997) find that $\sim$20\% of
the populations are extremely old and metal-poor. This appears to
contradict the observed positive correlation between FUV excess and
$Mg_2$ line strength (Burstein et al. 1988). Most attempts to explain the
FUV excess, using single stars, are based upon the idea of Greggio \& Renzini (1990)
that extremely low-mass metal-rich HB stars 
might skip the AGB phase altogether (``AGB-Manq\'ue stars'') and move directly to the EHB.

Brown et al. (1997), using Hopkins Ultraviolet Telescope (HUT) spectra
found that the spectral energy distributions of the UV excess in the cores of
bright ellipticals required both enhanced metallicity and enhanced
helium with respect to solar abundances. Dorman et al. (1995) present a range of spectral synthesis
models of old stellar populations in which the UV depends largely
on the distribution of envelope mass on the Zero Age Horizontal Branch. Yi et
al. (1997) present stellar population models with supersolar
metallicity and helium abundance, and a high, although ad-hoc, value of the Reimers
(1975) mass-loss parameter on the RGB ($\eta \sim 0.7 - 1.0$) and show
that they can reproduce the UV colours of ellipticals. 
Observations of a UV-bright Globular Cluster population in
M87 (Sohn et al. 2006, Kaviraj et al. 2007) again suggest a helium
enhancement. Chuzhoy \& Loeb (2004), Peng \& Nagai (2009) and others 
propose further that sedimentation of helium during the formation of
clusters and  galaxies could lead to enhanced helium in Brightest
Cluster Galaxies, which do indeed seem to show the strongest excess
(Ree et al 2007).  On the other hand Brown et al. (2008) present a
UV colour-magnitude diagram (CMD)  for M32, and note that it is
incompatible with their evolutionary tracks for supersolar helium
abundance. However evolutionary tracks on and beyond the RGB depend upon a
number of parameters, but most of all upon $\eta$, whose value, and
whose dependence upon other parameters of the population, are poorly
understood (Percival \& Salaris 2011). Moreover we are hampered by the
fact that M32, the one elliptical in which the UV bright stars can be
resolved, has a very weak UV upturn when compared with giant
ellipticals, and it is unclear that conclusions drawn from its CMD can
then be applied to more luminous galaxies. 

Evidence that the FUV excess is a phenomenon of the old stellar population is
provided by deep HST/STIS ultraviolet observations of ellipticals in the z=0.55
cluster CL0016+16 (Brown et al. 2000a), and the z=0.33 cluster
ZwCl1358.1+6245 (Brown et al. 2003), which show that the excess is far
weaker at these redshifts. The lookback times to these clusters, 3.7 and 5.4
Gyr respectively, suggest that the strengthening of the FUV occurs in populations older than 10 Gyr. 
However efforts to use the strength of the
excess as an age indicator are complicated by our paucity of 
observational evidence and lack of theoretical understanding of the dependence
of the phenomenon on other parameters of the stellar population (Brown et al. 2008).

Post AGB stars (central stars of Planetary Nebulae) have also
been proposed as a possible origin of the FUV excess,
but these are too hot to reproduce the observed spectral energy distribution, and
in addition observations of  M32 do not find them in
sufficient numbers (Brown et al. 2000b, 2008). 

Metal-poor, AGB-manque and binary star models make different predictions for the
dependence of the FUV excess upon stellar population age, metallicity
and $\alpha$-enhancement. The single-star models, predict different dependences
of the FUV excess on galaxy mass, metal abundance and population age
(Yi et al. 1999) whereas in the binary model the excess does not depend strongly on any
of these (Han et al. 2007; Lisker \& Han 2008). Specifically, in the
binary model, all old galaxies should show a UV excess at some level.

The GALaxy Evolution eXplorer (GALEX; Martin et al. 2005) has revolutionised the study of the global UV upturn in galaxies
and their associated star clusters. In galaxies it makes a systematic
study of the spatial distribution of the FUV excess stars possible. The
FUV excess has long been known to be more centrally concentrated in
ellipticals that the underlying old, red, population (O'Connell
1999). Gil de Paz et al. (2007) present ``The GALEX Ultraviolet atlas
of galaxies'', including surface brightness and colour profiles in the
GALEX far ultraviolet (FUV) and near ultraviolet (NUV) passbands for a
sample of 1034 nearby galaxies. Their (FUV-NUV) colour gradients of
bright early-type galaxies confirm the strong negative colour
gradients found in earlier data. They also find that elliptical and
lenticular galaxies with
brighter K-band absolute magnitudes show bluer (FUV-NUV) colours than
fainter ones. and they find that (FUV-K) is an excellent discriminant
of morphological type.

Donas et al. (2007), using the same data, investigate colour-colour
relations, and the relation between (FUV-NUV) and (B-V) colours and
the $Mg_2$ line index. They find a strong negative correlation between
the colours (galaxies which are red in B-V are blue in FUV-NUV), and a
negative correlation with the line index (galaxies with strong $Mg_2$
are bluer in (FUV-NUV). Replacing $Mg_2$ with the velocity dispersion,
$\sigma$, gives a very similar result, which is not surprising given
the well known correlation between $Mg_2$ and $\sigma$.

Loubser \& S\'anchez-Bl\'azquez (2011) analyse the global (FUV-NUV) colours of a sample
of Brightest Cluster galaxies (BCGs) and of a control sample of normal ellipticals, from the GALEX database. They 
examine the differences between BCGs and normal ellipticals, and the dependence of the strength of the global 
FUV excess upon stellar population parameters. They find that BGCs have stronger UV upturns than normal
ellipticals, and with less scatter between them. Surprisingly they find no significant dependence of the strength
of the FUV excess upon any stellar population parameter in the BCG sample. 

Marino et al. (2011) undertake surface photometry of a sample of gas-rich early type galaxies from GALEX images.
They confirm the relation between (FUV-NUV) colour and the
central velocity dispersion. Furthermore they investigate the relationship between the S\'ersic index $n$, derived in 
both UV and optical bands, and the stellar population parameters, and find that galaxies with $n > 4$ tend to have 
[$\alpha$/Fe] $>$ 0.15, and hence comparatively short star formation timescales.

Jeong et al. (2009) present GALEX surface photometry of a sample of galaxies which are also part of the SAURON survey. 
Their prime interest was in identifying early-type galaxies with residual star formation, however they do note that
their galaxies have predominantly positive (FUV-NUV) colour gradients, and that some of their sample have bluer
(FUV-NUV) at small radii than the global values for classic FUV excess galaxies. 

In this paper we use GALEX archive data, some of which is more recent
and deeper than the data available to Gil de Paz et al. (2007) and
Donas et al. (2007), to investigate the spatial distribution of the UV
excess, and to see how its properties depend upon other properties of
the galaxies. In Section~\ref{sec:Inputdata} we describe our sample
and briefly present the data used, which consists of archive data from
GALEX and the 2-Micron All-Sky Survey (2MASS). In
Subsection~\ref{sec:Apphot} we describe the measurement of annular
colours using circular apertures, and present colour gradients for our
sample. In Subsection~\ref{sec:Galfitphot}  we describe how we use the
galaxy surface brightness fitting programme \textsc{Galfit} (Peng et
al. 2002, 2010) to model the surface brightness profile in UV 
passbands, to provide more robust global colours and magnitudes than the GALEX
database values. In Subsection~\ref{sec:colourgradients} we examine the dependence of the colourgradients
upon other properties of the galaxies, and in Subsection~\ref{sec:stellarpops} we examine their dependence upon 
the stellar population parameters of the galaxies. 
In Subsection~\ref{sec:sersicindex} we investigate the correlation between the derived S\'ersic index and the 
stellar population parameters, suggested by Marino et al. (2011). In Subsection 
~\ref{sec:residualimages} we use the \textsc{Galfit} fits 
to the 2MASS and GALEX data to create residual images mapping the FUV excess, and we examine
and spatial extent of the UV excess upon other properties of the
galaxies. Finally in Section~\ref{sec:conclusions} we present our conclusions.

\section{Input Data}
\label{sec:Inputdata}

\subsection{GALEX UltraViolet data}

GALEX is a 50cm aperture space telescope optimised for imaging  in the range 1344 - 2832 {\AA}.
It also has a grism for low resolution spectroscopy, although we do not use any spectroscopic data in this 
study. The beam is split by a dichroic filter into two wavebands, 1344 - 1786 {\AA} (designated FUV) and 1771
- 2831 {\AA} (designated NUV). Details of the instrument, its mission and performance are given by 
Martin et al. (2005), Morrissey et al. (2005), and at the GALEX website\footnote{http://www.galex.caltech.edu/researcher/techdocs.html}. 

We recovered the majority of our data from the NASA MAST GR5 data archive\footnote{http://galex.stsci.edu/GR4/}. The exceptions were
NGC4486 (M87) for which the GR5 data were not available and for which we obtained the data later from the GR6 website\footnote{http://galex.stsci.edu/GR6/}, 
and the central Coma cluster galaxies NGC4874 and NGC4889, which were observed as part of programme GI5\_025 (PI: Smith). 

Our main sample consists of 29 passive E and cD type galaxies for which $V_T \lesssim 11.3$ and for which GALEX exposures of $\gtrsim$ 700 seconds
duration in the FUV band were available and the 3 brightest E/cD galaxies in Coma, NGC4874, NGC4839 and NGC4889, which are fainter than 
our nominal magnitude limit but for which very deep GALEX exposures exist. 

Although the conclusions of this paper are largely based upon this sample of 32 galaxies, we present and plot the results
for two separate and contrasting samples for comparison. The first is a sample of 10 E and cD galaxies for which there is evidence
some evidence of nuclear activity, as noted in the NASA/IPAC Extragalactic Database (NED)\footnote{http://nedwww.ipac.caltech.edu/}.
These galaxies range from the very strongly active NGC4486, to weak liners and low-luminosity AGN, including
NGC4374 and NGC 4552. The other is a sample of 10 galaxies with elliptical morphology but which are given the type S0, SA0 or SAB0
in RC3 (De Vaucouleurs et al. 1991) and therefore in NED. The reasons for this classification are either the presence of dust lanes,
shell structure, or star formation regions (or in some cases all three). 
We keep the three samples separate in our analysis.

\begin{table*}
\begin{tabular}{llccccc}
Galaxy&Type&Distance&$V_T^0$&GALEX Dataset&NUV&FUV\\
&&Modulus&&&exposure (s)&exposure(s)\\ \hline
\multicolumn{7}{l}{Passive E and cD galaxies}\\ \hline
NGC 584&E4&31.44&10.30&NGA\_NGC0584&1697&1697\\
NGC 596&cDpec&31.56&10.80&NGA\_NGC0584&1697&1697\\
NGC 720&E5&31.88&10.17&HRC\_RXJ0152m1357&57037&55734 \\
NGC 1374&E&31.42&11.10&UVE\_FORNAX&34815&33787 \\
NGC 1379&E&31.29&10.99&UVE\_FORNAX&34815&33787 \\
NGC 1399&cD&31.36&9.49&UVE\_FORNAX&34815&33787 \\
NGC 1404&E1&31.33&10.03&UVE\_FORNAX&34815&33787 \\
NGC 1407&E0&31.85&9.74&NGA\_NGC1407&1557&1557 \\
NGC 1549&E0-1&31.04&9.76&NGC\_NGC1546&2011&2011\\
NGC 3258&E1&33.01&11.30&GI3\_087006\_NGC3268&2223&2223\\
NGC 3268&E2&33.02&11.30&GI3\_087006\_NGC3268&2223&2223\\
NGC 3377&E5-6&30.14&10.23&GI3\_084014\_J104728p135322&1645&1645\\
NGC 3379&E1&30.36&9.24&GI4\_042055\_J104657p125223&1153&1153\\
NGC 3608&E2&31.87&10.76&GI3\_079016\_NGC3608&2468&2468\\
NGC 3923&E4-5&31.57&9.69&NGA\_NGC3923&3019&2269\\
NGC 3962&E1&32.38&10.67&GI3\_087006\_NGC3962&2098&2098\\
NGC 4365&E3&31.63&9.54&GI2\_125012\_AGESstrip1\_12&4523&1580\\
NGC 4406&E3&31.04&8.84&NGA\_VIRGO\_MOS10&3108&1580\\
NGC 4473&E5&31.00&10.11&NGA\_VIRGO\_MOS09&4521&1387\\
NGC 4621&E5&31.01&9.61&GI3\_041008\_NGC4621&1658&1658\\
NGC 4649&E2&31.09&8.75&GI3\_041008\_NGC4621&1658&1658\\
NGC 4697&E6&30.25&9.18&GI4\_085003\_NGC4697&1696&1696\\
NGC 4839&cD&34.88&11.95&GI2\_046001\_COMA3&31165&29997\\
NGC 4874&cD0&34.97&11.73&GI5\_025001\_COMA&14692&13588\\
NGC 4889&E4&34.88&11.37&GI5\_025001\_COMA&14692&13588\\
NGC 5044&E0&32.64&11.03&GI3\_087011\_NGC5044&1696&1696\\
NGC 5813&E1-2&32.39&10.48&NGA\_NGC5813&693&693\\
NGC 5831&E3&32.22&11.39&GI1\_109008\_NGC5831&5325&2368\\
NGC 5846&E0-1&32.13&9.95&NGA\_NGC5846&1117&1117\\
NGC 5982&E3&33.07&11.17&GI1\_109009\_NGC5982&3586&1567\\
NGC 6868&E2&32.41&10.54&ABELL\_S0851\_NGC6868&6417&2375\\
NGC 6958&cD&32.59&11.28&NGA\_NGC6958&3094&3094\\
&&&&&&\\ \hline
\multicolumn{7}{l}{Liners and other weak AGN}\\ \hline
NGC 1052&E4&31.46&10.44&NGC\_NGC1052&3800&2967\\
NGC 2768&E6&31.49&9.78&NGA\_NGC2768&1647&1647\\
NGC 4261&E2-3&32.53&10.39&GI3\_079021\_NGC4261&1645&1645\\
NGC 4278&E1-2&30.85&10.07&NGC\_NGC4278&1481&1481\\
NGC 4374&E1&31.19&9.07&NGA\_VIRGO\_MOS10&3108&1580\\
NGC 4486&cDpec&31.10&8.56&GI1\_077011\_TYC8775461&1702&1702\\
NGC 4552&E0-1&30.96&9.63&NGA\_VIRGO\_MOS03&4762&1598\\
NGC 4589&E2&32.27&10.75&GI3\_079022\_NGC4589&3400&3399\\
IC 1459&E3-4&32.14&9.88&GI1\_093001\_IC1459&1677&1677\\ 
IC 4296&E0&33.46&10.47&GI3\_087015\_IC4296&3365&1701\\
&&&&&&\\ \hline
\multicolumn{7}{l}{Starforming galaxies and early-type spirals}\\ \hline
NGC 474&SA0&32.44&11.15&NGRG\_A227&2708&2708\\
NGC 1316&SAB0pec&31.41&8.53&NGA\_NGC1316&1702&1702 \\
NGC 1387&SAB0&31.26&10.72&UVE\_FORNAX&34815&33787 \\
NGC 1389&SAB0&31.16&11.54&UVE\_FORNAX&34815&33787 \\
NGC 1400&SA0&31.73&10.97&NGA\_NGC1407&1557&1557 \\
NGC 1553&SA0&30.68&9.39&NGA\_NGC1546&2011&2011\\
NGC 2865&E3-4&32.40&11.36&GI1\_059003\_NGC2865&16251&2560\\
NGC 3115&S0&29.99&8.80&NGA\_NGC3115&1304&1304\\
NGC 3384&SB0&30.29&9.84&GI4\_042055\_J104657p125223&1153&1153\\
NGC 4459&SA0&31.11&10.29&GI1\_109010\_NGC4459&1593&1593\\
\hline
\end{tabular}
\caption{\label{tab:GALEXdata}Galaxies observed and the properties of the images used. Columns 2, 3 and 4
give the type, distance and total V band magnitude from the NASA Extragalactic Database (NED). Column 5 gives
the name of the GALEX dataset used, in some cases the primary target was not the galaxy of our study. Columns
6 and 7 list the exposure time in seconds for the NUV and FUV images respectively.}
\end{table*}

Our reductions all assume the
GR5 zero points for the data, which are 18.82 mag in FUV and 20.08 mag in NUV. No linearity corrections were applied
to the data, and our error analyses include only Poisson noise from sources and background (whether sky or detector
background). The pixel size is 1.5 arcsec square. 

\subsection{InfraRed data}

InfraRed data for the galaxies listed in Table~\ref{tab:GALEXdata} was obtained from the Interactive 2MASS  (Skrutskie et al. 2006)  image 
server\footnote{http://irsa.ipac.caltech.edu/applications/2MASS/IM/interactive.html}. 
2MASS Atlas images, rather than Quicklook images were used for our photometric analysis. 
However for the {\tt Galfit} analysis described in Section ~\ref{sec:Galfitphot}, the galaxies were often close to the edges of the 
frames, which compromised the fits. For the larger galaxies we therefore used images from the 2MASS Large Galaxy Atlas
(Jarrett et al. 2003) for the {\tt Galfit} fits. 
2MASS Atlas images are presented with a pixel scale of 1 arcsec/pixel, but are resampled from 2 arcsec pixel scans. 

\section{Results}

Our results, and properties determined from external sources, are summarised in Table~\ref{tab:properties}. In this Table,
data presented in columns 3-8 are determined in this study; the absolute magnitude given in column 2 comes from the 
extinction corrected apparant V-band magnitude and distance modulus given in NED; the velocity dispersion  comes from 
various sources listed in \textsl{Hyperleda}\footnote{http://leda.univ-lyon1.fr/} as described in the text, and the stellar 
population parameters given in columns 9-11 come from the source indicated in column 12, the key to the references is given
in the table caption.

\subsection{Aperture photometry}
\label{sec:Apphot}

Aperture photometry was carried out in a range of circular apertures, of radius 3.0, 5.0, 8.0, 10.0, 15.0, 20.0, 30.0, 40.0 and 55.0
arcsec, using the {\textsc IRAF} task {\textsc PHOT}. As described by
Carter et al. (2009), we determined the sky background in a number (8-12) of circular apertures located away from the galaxies 
and in regions free from stars. Following Gil de Paz et al. (2007), we use an unclipped mean as the background measurement for the UV images,
as the statistics are poissonian and in many pixels in the FUV images the background is zero. 
The centroid of the galaxy was determined within {\textsc PHOT}. Surface brightnesses within circular annuli were determined from the 
differences between intensities within apertures. Intensity and magnitude  errors were calculated as described by Carter et al. (2009). 
Statistical errors from the image and background were taken from count statistics alone in the case of GALEX, and calculated 
according to the prescription presented on the 2MASS website\footnote{http://www.ipac.caltech.edu/2mass/releases/allsky/doc/sec6\_8a.html}
in the case of 2MASS. To these were added a measurement repeatability error determined from four separate measurements. As the main
focus of this paper is colour gradients and differences, no zero point errors were included.  Surface brightness
errors were obtained by adding in quadrature the intensity errors derived from two apertures which were subtracted. 

Foreground stars contaminated a small number of annuli, the intensities of these were measured in 3 arcsec radius apertures, for both GALEX and 2MASS
images, and they were subtracted. No corrections were made for the ellipticity of the galaxies, or for the different point spread functions of the images. 
Corrections for galactic extinction are taken from the values of Schlegel et al. (1998) 
as tabulated in NED\footnote{http://nedwww.ipac.caltech.edu}, and from the relationships between E(B-V) and the extinction 
in the GALEX bands given by Wyder et al. (2005). 

\subsubsection{The effect of the Point Spread Function}

\begin{figure}
\begin{center}
\includegraphics[width=8.5cm]{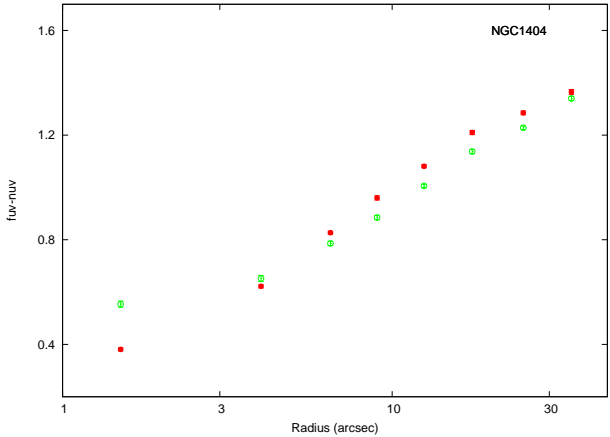}
\caption{(FUV-NUV) plotted against radius for NGC1404. The filled circles show the colour values derived from the unconvolved
GALEX images, and the open symbols show those derived from the ``PSF matched'' images, where the FUV image is convolved
with the NUV PSF and vice-versa. The difference in the PSFs affects mainly the central circular aperture, which is omitted when the 
fits to determine the logarithmic gradients are made.
}
\label{fig:psfmatch}
\end{center}
\end{figure}

The Point Spread Function (PSF) differs between the two GALEX bands, and in the 2MASS bands it varies from image to image. In order
to assess the effect of differences in the PSF on the (FUV-NUV) gradient in particular, we  test the effect of matching the resolutions by
convolving the Fornax cluster FUV image with the NUV PSF, and the NUV image with the FUV psf. In Figure~\ref{fig:psfmatch} we show
the (FUV-NUV) colour profile of NGC1404 as determined from the unconvolved images (open circles) and the images convolved to
match the psf (filled squares). Based upon this plot, for each galaxy we measure the logarithmic colour gradients, $\nabla_{FN}$ and $\nabla_{NJ}$, which are the 
the slopes of the least squares regression lines of (FUV-NUV) and (NUV-J) repectively upon $\log{radius}$, but in each fit we omit the colour in the
central circular aperture. Besides the effect of the different PSFs on this aperture, it is potentially affected by non-stellar emission from the nucleus.
Apart from omitting the central aperture we make no correction for the differing PSFs.  These gradients are 
presented in Table~\ref{tab:properties} and discussed in Section~\ref{sec:colourgradients}. 

In Figures A1 to A13, of Appendix A, presented in the online version of the paper only, we 
show surface brightness and colour profiles for our samples. All photometry is corrected for galactic extinction as discussed above.
All galaxies apart from NGC2865 show positive gradients in (FUV-NUV). Also plotted on the  individual panels of 
the figures in the online version of Appendix A are the fits, showing the gradient from Table~\ref{tab:properties}.

\begin{table*}
{\scriptsize
\begin{tabular}{lcccccccccccc}
Galaxy&$-M_V^0$&N$^0$&(F-N)$^0_G$&(F-N)$^0_{{R_e}/8}$&(N-H)$^0$&$\nabla_{FN}$&$\nabla_{NJ}$&Log$\sigma_0$&[Z/H]&[$\alpha$/Fe]&Ref\\ \hline
&&&&&&\\
NGC 584&21.14&15.75&2.44&1.76$\pm$0.08&6.90&0.34$\pm$0.14&-0.59$\pm$0.04&2.330$\pm$0.015&0.35$\pm$0.09&0.16$\pm$0.02&1,6\\
NGC 596&20.76&15.96&2.62&1.60$\pm$0.07&6.57&0.72$\pm$0.15&-0.55$\pm$0.09&2.209$\pm$0.011&0.19$\pm$0.07&0.09$\pm$0.03&1,5,6\\
NGC 720&21.71&15.44&1.31&0.49$\pm$0.01&6.81&1.10$\pm$0.04&-0.26$\pm$0.04&2.393$\pm$0.009&0.39$\pm$0.07&0.34$\pm$0.02&1,6\\
NGC 1374&20.32&15.95&1.70&0.98$\pm$0.03&6.37&0.60$\pm$0.10&-0.76$\pm$0.04&2.238$\pm$0.011&&&\\
NGC 1379&20.30&15.94&1.85&1.38$\pm$0.01&6.78&0.41$\pm$0.07&-0.59$\pm$0.03&2.090$\pm$0.013&-0.13$\pm$0.05&0.30$\pm$0.01&1\\
NGC 1399&21.87&14.59&0.54&0.07$\pm$0.03&6.77&0.68$\pm$0.07&0.25$\pm$0.04&2.525$\pm$0.009&0.44$\pm$0.06&0.40$\pm$0.03&1,7\\
NGC 1404&21.30&15.22&1.28&0.49$\pm$0.01&6.94&0.66$\pm$0.09&-0.21$\pm$0.06&2.378$\pm$0.008&&&\\
NGC 1407&22.11&15.06&1.31&0.41$\pm$0.02&6.99&0.60$\pm$0.02&0.03$\pm$0.06&2.452$\pm$0.008&0.36$\pm$0.03&0.29$\pm$0.02&1,2,4,6\\
NGC 1549&21.28&14.68&1.73&1.14$\pm$0.03&6.34&0.63$\pm$0.07&-0.50$\pm$0.03&2.317$\pm$0.004&0.28$\pm$0.04&0.20$\pm$0.01&1,6\\
NGC 3258&21.71&16.71&0.96&0.40$\pm$0.04&7.20&1.27$\pm$0.09&0.25$\pm$0.07&2.438$\pm$0.019&0.42$\pm$0.11&0.21$\pm$0.03&2\\
NGC 3268&21.72&16.82&1.33&0.68$\pm$0.05&7.24&0.87$\pm$0.17&-0.58$\pm$0.08&2.368$\pm$0.008&0.11$\pm$0.07&0.34$\pm$0.04&2\\
NGC 3377&19.91&14.96&2.99&1.85$\pm$0.05&5.91&0.21$\pm$0.11&-0.96$\pm$0.04&2.162$\pm$0.009&0.24$\pm$0.02&0.24$\pm$0.02&3,4,6\\
NGC 3379&21.12&14.79&2.00&1.06$\pm$0.04&7.02&0.51$\pm$0.03&-0.52$\pm$0.02&2.336$\pm$0.005&0.19$\pm$0.02&0.20$\pm$0.02&3,4,6\\
NGC 3608&21.13&16.19&1.99&0.81$\pm$0.03&6.87&0.86$\pm$0.06&-0.54$\pm$0.06&2.306$\pm$0.010&0.29$\pm$0.02&0.25$\pm$0.02&3,4,6\\
NGC 3923&21.88&14.62&2.03&0.65$\pm$0.02&6.70&0.86$\pm$0.08&-0.20$\pm$0.03&2.387$\pm$0.015&0.54$\pm$0.08&0.27$\pm$0.01&1,6\\
NGC 3962&21.71&15.72&2.15&1.05$\pm$0.04&6.62&0.63$\pm$0.11&-0.55$\pm$0.05&2.359$\pm$0.012&0.13$\pm$0.02&0.22$\pm$0.03&2\\
NGC 4365&22.09&14.35&2.07&0.44$\pm$0.02&6.35&0.91$\pm$0.08&-0.05$\pm$0.06&2.420$\pm$0.005&0.19$\pm$0.01&0.28$\pm$0.01&4\\
NGC 4406&22.20&14.13&2.18&1.04$\pm$0.02&6.44&0.47$\pm$0.07&-0.41$\pm$0.02&2.396$\pm$0.006&0.40$\pm$0.03&0.23$\pm$0.01&5\\
NGC 4473&20.89&15.36&2.08&1.08$\pm$0.05&6.62&0.52$\pm$0.15&-0.74$\pm$0.01&2.270$\pm$0.007&0.30$\pm$0.02&0.24$\pm$0.04&3\\
NGC 4621&21.40&15.25&1.83&0.56$\pm$0.02&6.99&0.80$\pm$0.11&-0.23$\pm$0.07&2.381$\pm$0.008&0.30$\pm$0.04&0.28$\pm$0.05&3\\
NGC 4649&22.34&14.14&1.21&0.34$\pm$0.01&7.08&0.77$\pm$0.04&0.36$\pm$0.06&2.544$\pm$0.005&0.36$\pm$0.03&0.30$\pm$0.01&6\\
NGC 4697&21.07&14.24&2.45&1.37$\pm$0.03&6.46&0.43$\pm$0.06&-0.50$\pm$0.03&2.242$\pm$0.005&0.15$\pm$0.04&0.14$\pm$0.02&1,2,4,6\\
NGC 4839&22.93&17.43&1.24&0.51$\pm$0.01&6.93&0.88$\pm$0.10&-0.18$\pm$0.11&2.430$\pm$0.008&0.15$\pm$0.03&0.33$\pm$0.02&6,7\\
NGC 4874&23.24&16.54&1.67&0.77$\pm$0.03&6.70&0.49$\pm$0.04&-0.61$\pm$0.04&2.419$\pm$0.009&0.53$\pm$0.03&0.41$\pm$0.03&6,7\\
NGC 4889&23.51&16.74&1.18&0.45$\pm$0.02&7.01&0.77$\pm$0.02&-0.03$\pm$0.08&2.593$\pm$0.005&0.67$\pm$0.03&0.40$\pm$0.02&6,7\\
NGC 5044&21.61&15.32&1.48&0.59$\pm$0.03&6.61&0.62$\pm$0.06&-0.43$\pm$0.05&2.380$\pm$0.018&0.12$\pm$0.20&0.30$\pm$0.09&2,4\\
NGC 5813&21.91&15.94&2.05&0.83$\pm$0.06&7.04&0.70$\pm$0.21&-0.56$\pm$0.14&2.380$\pm$0.006&0.11$\pm$0.03&0.29$\pm$0.02&2,3,6\\
NGC 5831&20.83&16.25&2.71&1.06$\pm$0.06&6.37&0.82$\pm$0.17&-0.52$\pm$0.03&2.224$\pm$0.011&0.28$\pm$0.14&0.16$\pm$0.03&1,2,3,6\\
NGC 5846&22.18&15.31&1.27&0.58$\pm$0.03&7.33&0.41$\pm$0.09&-0.02$\pm$0.10&2.409$\pm$0.009&0.26$\pm$0.03&0.27$\pm$0.02&2,3,6\\
NGC 5982&21.90&16.32&1.93&0.63$\pm$0.05&6.69&0.95$\pm$0.14&-0.03$\pm$0.08&2.408$\pm$0.007&0.38$\pm$0.02&0.28$\pm$0.04&3\\
NGC 6868&21.87&15.90&1.66&0.84$\pm$0.04&7.25&0.46$\pm$0.12&-0.24$\pm$0.07&2.421$\pm$0.007&0.26$\pm$0.07&0.19$\pm$0.03&2\\
NGC 6958&21.31&16.37&2.43&1.44$\pm$0.06&6.60&0.82$\pm$0.04&-0.63$\pm$0.11&2.293$\pm$0.012&0.11$\pm$0.06&0.15$\pm$0.02&1,3,6\\
&&&&&&&&&&&\\ \hline
NGC 1052&21.02&15.77&1.50&0.61$\pm$0.02&7.03&0.89$\pm$0.07&-0.17$\pm$0.11&2.330$\pm$0.009&0.24$\pm$0.05&0.37$\pm$0.03&2,6\\
NGC 2768&21.71&15.13&2.17&1.27$\pm$0.04&6.85&0.39$\pm$0.08&-0.63$\pm$0.04&2.294$\pm$0.008&0.22$\pm$0.04&0.26$\pm$0.04&3\\
NGC 4261&22.14&15.80&1.28&0.31$\pm$0.02&6.84&0.92$\pm$0.04&-0.05$\pm$0.10&2.496$\pm$0.005&0.28$\pm$0.04&0.26$\pm$0.02&6\\
NGC 4278&20.78&15.34&1.41&0.37$\pm$0.02&6.69&0.81$\pm$0.06&-0.20$\pm$0.19&2.411$\pm$0.007&0.22$\pm$0.02&0.40$\pm$0.03&3,6\\
NGC 4374&22.14&14.47&1.87&0.91$\pm$0.02&6.71&0.65$\pm$0.05&-0.31$\pm$0.05&2.470$\pm$0.004&0.23$\pm$0.06&0.27$\pm$0.02&2,3,4,6\\
NGC 4486&22.54&13.96&1.11&0.47$\pm$0.01&6.82&0.30$\pm$0.06&0.54$\pm$0.10&2.541$\pm$0.007&0.34$\pm$0.02&0.41$\pm$0.05&3\\
NGC 4552&21.33&15.20&1.07&0.18$\pm$0.02&6.90&0.80$\pm$0.08&0.19$\pm$0.11&2.427$\pm$0.005&0.37$\pm$0.04&0.17$\pm$0.02&2,3,4,6\\
NGC 4589&21.52&15.76&2.47&1.40$\pm$0.04&6.51&0.58$\pm$0.04&-0.95$\pm$0.06&2.353$\pm$0.011&&&\\
IC 1459&22.26&14.57&2.05&0.43$\pm$0.02&6.09&0.91$\pm$0.09&-0.08$\pm$0.06&2.497$\pm$0.006&0.37$\pm$0.08&0.25$\pm$0.04&2\\ 
IC 4296&22.99&15.71&1.87&0.82$\pm$0.04&6.77&0.81$\pm$0.11&0.04$\pm$0.06&2.519$\pm$0.008&0.42$\pm$0.12&0.27$\pm$0.02&1,2\\
&&&&&&&&&&&\\ \hline
NGC 474&21.29&16.19&2.89&1.56$\pm$0.06&5.86&0.73$\pm$0.12&-1.03$\pm$0.17&2.231$\pm$0.013&0.26$\pm$0.04&0.16$\pm$0.05&3\\
NGC 1316&22.88&13.44&2.29&1.57$\pm$0.02&6.29&0.16$\pm$0.09&-0.59$\pm$0.04&2.366$\pm$0.009&0.34$\pm$0.02&0.15$\pm$0.01&6\\
NGC 1387&20.54&15.76&0.96&0.52$\pm$0.01&6.70&0.92$\pm$0.08&0.91$\pm$0.33&2.231$\pm$0.029&&&\\
NGC 1389&19.62&16.61&2.08&1.53$\pm$0.05&6.64&0.49$\pm$0.06&-0.52$\pm$0.05&2.112$\pm$0.028&0.25$\pm$0.06&0.08$\pm$0.03&2\\
NGC 1400&20.76&15.79&1.08&0.55$\pm$0.04&6.50&0.91$\pm$0.16&0.51$\pm$0.26&2.412$\pm$0.004&0.37$\pm$0.06&0.27$\pm$0.01&1,4,6\\
NGC 1553&21.29&14.27&2.32&1.30$\pm$0.02&6.71&0.67$\pm$0.08&-0.41$\pm$0.03&2.262$\pm$0.013&0.35$\pm$0.03&0.15$\pm$0.01&1,3,6\\
NGC 2865&21.04&16.12&2.35&2.62$\pm$0.13&6.12&-0.36$\pm$0.16&-0.03$\pm$0.16&2.251$\pm$0.010&0.10$\pm$0.02&0.36$\pm$0.03&4\\
NGC 3115&21.19&14.37&1.89&0.68$\pm$0.02&6.77&0.60$\pm$0.10&-0.29$\pm$0.04&2.435$\pm$0.007&0.61$\pm$0.10&0.20$\pm$0.03&1\\
NGC 3384&20.45&15.25&2.75&1.40$\pm$0.06&6.74&0.44$\pm$0.17&-0.78$\pm$0.08&2.196$\pm$0.011&0.37$\pm$0.01&0.22$\pm$0.02&3,4\\
NGC 4459&20.82&15.44&2.19&1.11$\pm$0.03&6.94&1.19$\pm$0.12&0.75$\pm$0.20&2.256$\pm$0.021&0.38$\pm$0.02&0.22$\pm$0.04&3\\
\hline
\end{tabular}}
\caption{\label{tab:properties}Ultraviolet and stellar population properties of the sample. Column 2 gives the absolute magnitude (from the 
apparent magnitude and distance modulus given in Table ~\ref{tab:GALEXdata}). Columns 3, 4 and  6 give the extinction corrected apparent 
NUV magnitude, (FUV-NUV) and (NUV-H) colours, derived with \textsc{Galfit}. Column 5 gives extinction corrected (FUV-NUV) within a central aperture
of radius $R_e/8$
as measured with \textsc{PHOT}. Columns 7 and 8 give the colour gradients in (FUV-NUV) and (NUV-J), derived
from the data presented in Appendix A in the online version of the paper, omitting the central value. Column 9 gives the logarithm of the central stellar velocity 
dispersion $\sigma_0$, derived from the values listed in \textit{Hyperleda}. Columns 10-12 give stellar population parameters (Log age, [Z/H], [$\alpha$/Fe]. Column
12 gives the  source of the stellar population paremeters, coded as follows: 1=Ogando et al. (2008, 2010); 2=Annibali et al. (2007), their values of Z converted
to [Z/H] using Z$_{\odot}$ = 0.018; 3=Kuntschner et al. (2010);
4=Spolaor at el. (2010); 5=Serra et al. (2008); 6=Thomas et al. (2005); 7=Loubser et al. (2009)}
\end{table*}

\subsection{Global Colours}
\label{sec:Galfitphot}

Using \textsc{Galfit} (Peng et al. 2002,2010) we fit S\'ersic (1962, 1968) profiles to the FUV, NUV and H-band images. From these fits
we derive the S\'ersic index $n$ in all three bands, and the total magnitudes in the GALEX bands. Because of the strong coupling between 
the three parameters of the fit ($n$, effective radius and total magnitude),  we present total magnitudes in the GALEX bands constrained
to $n$ = 4, when the S\'ersic profile reduces to the De Vaucouleurs (1948) form. Although this runs the risk of underestimating the light 
at large radii, in this paper we are interested only in the (FUV-NUV) colour, and we believe that the colours derived by using \textsc{Galfit} in this 
way will be more precise than the \textsc{SExtractor}-derived GALEX calatogue values, and than the ``Asymptotic'' colours derived from
growth curves by Gil de Paz et al. (2007), which are in any case only available for a subset of our sample. 

\begin{figure}
\begin{center}
\includegraphics[width=8cm]{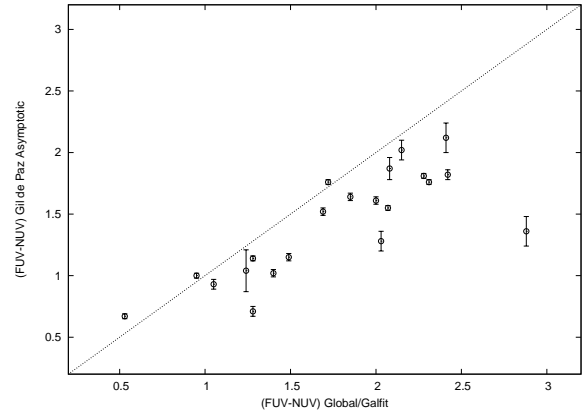}
\caption{Comparison between global (FUV-NUV) as derived from our \textsc{Galfit} fits, and the ``Asymptotic''
colours of Gil de Paz et al. (2007). The Gil de Paz et al. colours are $\approx$ 0.25 mag bluer than  the \textsc{Galfit} 
colours, suggesting that \textsc{Galfit} is sampling the galaxy to larger radius than the aperture magnitudes from which the
Asymptotic magnitude is derived.
}
\label{fig:comparison}
\end{center}
\end{figure}

In Figure~\ref{fig:comparison} we compare our (FUV-NUV) colours with those from Gil de Paz et al. (2007). Their Asymptotic colours
are in most cases bluer than our global colours, the offset is around 0.25 magnitude. Rather than being due to systematic errors in 
either technique, this difference is due to the fact that \textsc{Galfit} effectively extrapolates to larger radii than those of the the apertures used to
define the growth curve, and thus the outer, redder regions are included in the \textsc{Galfit} colours.
In this paper we use the \textsc{Galfit}-derived magnitudes, and these magnitudes and colours are presented in Table~\ref{tab:properties}.

\begin{figure}
\begin{center}
\includegraphics[width=8.5cm]{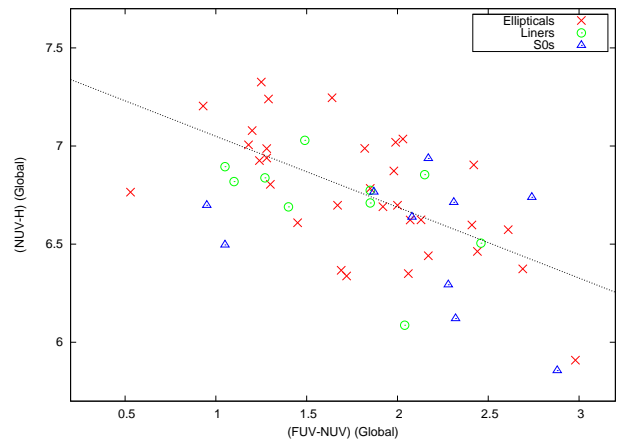}
\caption{Colour-colour diagram, showing global (NUV-H) against (FUV-NUV).  In this and subsequent
plots the red crosses represent the normal ellipticals, the green open circles the ellipticals with Liner activity,
and the blue triangles the sample of star forming ellipticals and SA/S0 galaxies. The ellipticals show a clear
trend, galaxies which are bluer in (FUV-NUV) are redder in (NUV-H). 
}
\label{fig:colour-colour}
\end{center}
\end{figure}

In Figure~\ref{fig:colour-colour} we present a colour-colour diagram, plotting the global IR-UV colour (NUV-H) against 
(FUV-NUV). There is a weak but clear negative correlation, for the 32 ellipticals (i.e. the red crosses in the Figure)  the Spearman 
Rank Correlation Coefficient (SRCC)  is --0.62, giving significance level $p = 0.00014$. The NUV magnitude appears on both 
axes in this figure, but the measurement errors on NUV are small and do not contribute significantly to the correlation.

\begin{figure*}
\begin{center}
\includegraphics[width=16cm,trim=0cm 0cm 0cm 3.5cm]{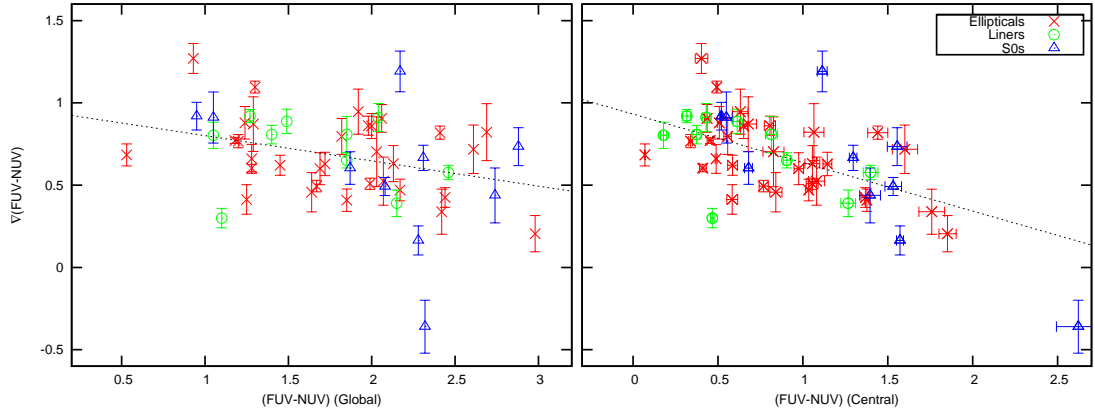}
\caption{Logarithmic gradient in (FUV-NUV) plotted against global (FUV-NUV)  in the left panel
and central (FUV-NUV) in the right panel. In each panel the ellipticals show a weak
but clear trend, the strongest gradients are in the galaxies with the bluest (FUV-NUV) colours. The 
dotted line represents a least-squares fit to the data.
}
\label{fig:grad-colour}
\end{center}
\end{figure*}

In Figure ~\ref{fig:grad-colour} we plot the logarithmic radial gradient in (FUV-NUV) against its global value. 
There is a weak negative trend in this plot, but it is not statistically significant (SRCC = --0.26; p = 0.15  for the 32 ellipticals). 
The Pearson correlation coefficient for this particular correlation is --0.373, giving $p \sim 0.03$, but this is largely driven 
by two points, and in this case we regard the Spearman test as more robust. 

In the right panel of this figure we plot for comparison the logarithmic radial gradient against the central value
of (FUV-NUV), that within a central aperture of radius $R_e/8$, $R_e$ being the effective radius as taken from De Vaucouleurs et al. (1991).
$R_e/8$ lies in the range 1.5 - 13.0 arcsec, and as the PSFs in FUV and NUV are different, there will be a systematic
offset in (FUV-NUV) which depends upon its value. From measurement of aperture colours of a model galaxy image
convolved with the two PSFs, we determine and apply aperture corrections for the central  colours. These corrections
are in the range 0.05 -- 0.11 magnitude.

Here the correlation appears more significant (SRCC = -- 0.51; p = 0.003),
although this is not surprising, the strongest gradients are in those galaxies with the bluest central colours

\begin{figure*}
\begin{center}
\includegraphics[width=16cm]{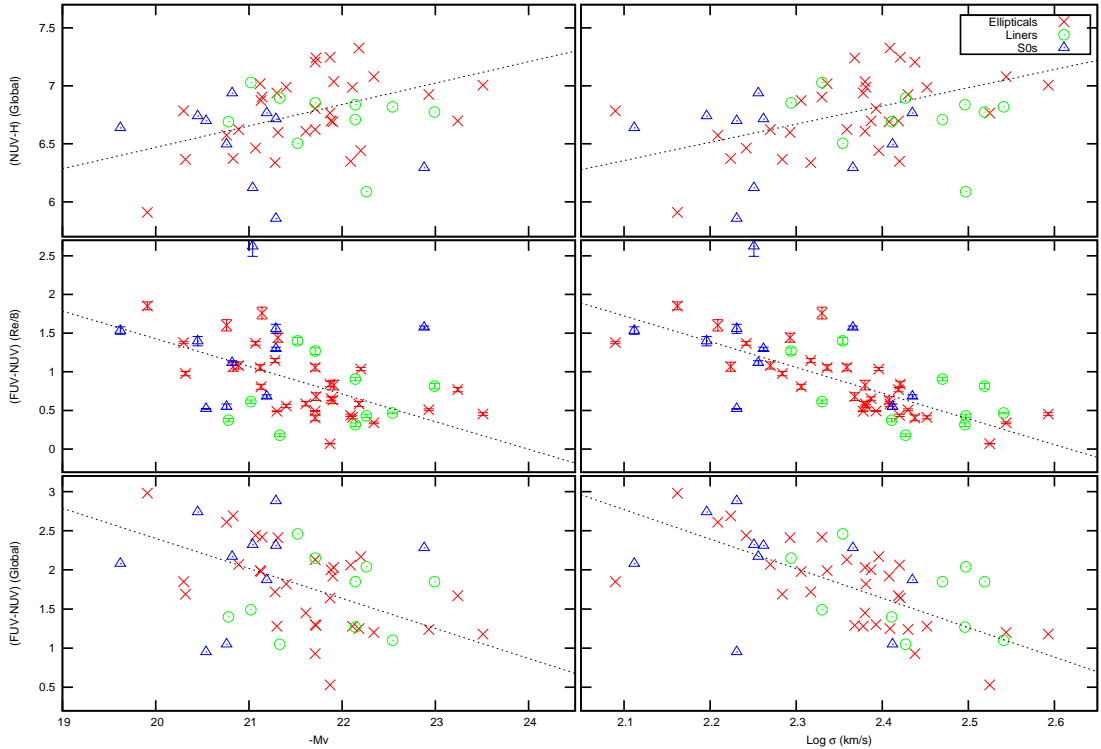}
\caption{Dependence of global UV (lower panels), central UV (centre panels) and UV-IR (upper panels) colours upon absolute V-band magnitude (left)
and central velocity dispersion (right). The ellipticals show a particularly strong trend with both global and central (FUV-NUV), with the high
velocity dispersion galaxies being bluer. The dashed lines represent least squares fits, the significance of the correlations is discussed in the text.
\label{fig:ColourPlots}
}
\end{center}
\end{figure*}

Figure~\ref{fig:ColourPlots} shows the global colours plotted against V-band absolute magnitude ($M_V$) and the logarithm of central velocity dispersion ($\sigma_0$). 
We take $\sigma_0$ from the list of measurements given in \textsl{HyperLeda}. This lists measurements
of $\sigma_0$ from a number of sources, we ignore those which are clearly not central values (e.g. measurements of $\sigma$ in a system 
of Planetary Nebulae) or which are taken from very old (e.g. photographic) spectra. Then for those galaxies with five measurements or fewer we 
take the mean, for those with more than five we reject the highest and the lowest and take the mean of the remainder. Our 
adopted velocity dispersions are given as $\log{\sigma_0}$ in Table~\ref{tab:properties}, where we also give an error, which for galaxies with $>$5 measurements
is the standard error of the mean. Both colours show significant correlations with both $-M_V$ and $\log{\sigma_0}$, in the sense that brighter galaxies
and those with high $\log{\sigma_0}$ are bluer in (FUV-NUV) and redder in (NUV-H). In agreement with Marino et al. (2011) we find
the most significant correlation to be that between (FUV-NUV) and $\log{\sigma_0}$ (SRCC = --0.74; $p < 0.0001$ for the ellipticals), suggesting that 
the depth of the potential well, more than the total mass, plays a key r\^ole in determining the strength of the UV excess. A multiple regression upon
$-M_V$ and $\log{\sigma_0}$ confirms that all of the trend is accounted for by  $\log{\sigma_0}$,
with no significant residual luminosity dependence. The correlations with central (FUV-NUV) are slightly stronger (SRCC = -0.86; $p = 10^{-9}$ for the 
$\log{\sigma_0}$ case) which is unsurprising as $\log{\sigma_0}$ is a central measurement.

\subsection{Colour Gradients}
\label{sec:colourgradients}

\begin{figure*}
\begin{center}
\includegraphics[width=16cm]{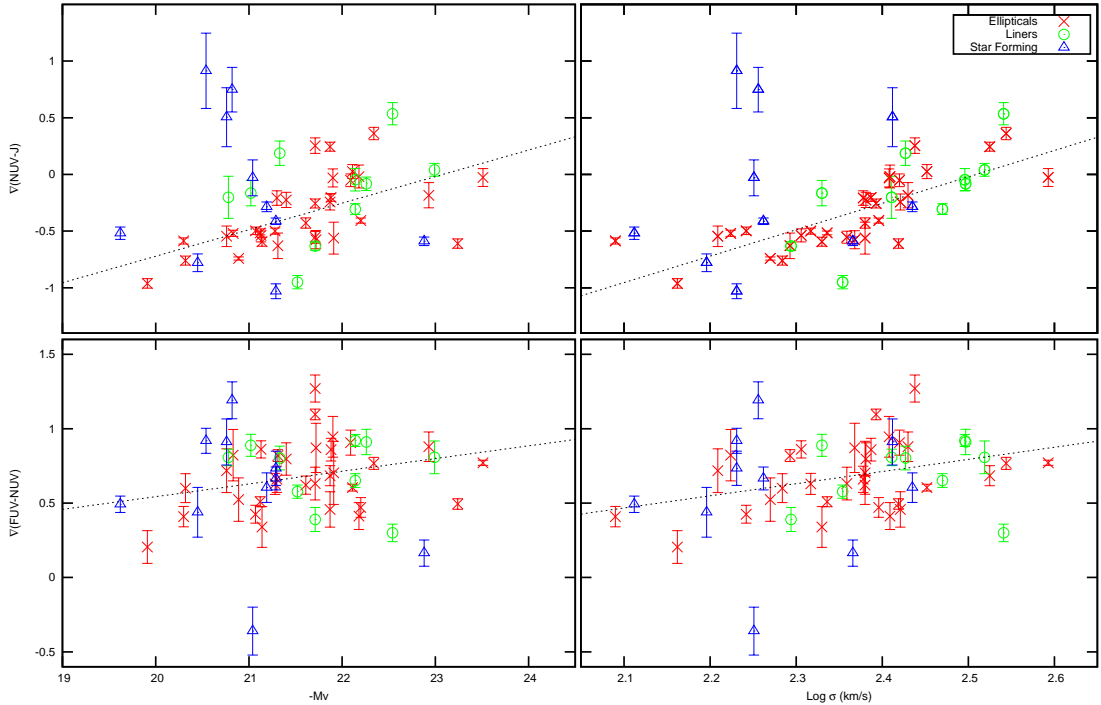}
\caption{Dependence of UV (lower panels) and UV-IR (upper panels) colour gradients upon absolute V-band magnitude (left)
and central velocity dispersion (right). All galaxies except one show positive gradients in (FUV-NUV). There are strong 
negative gradients in (NUV-J), particularly in the lower velocity dispersion ellipticals. The dashed lines represent least squares fits, 
the significance of the correlations is discussed in the text.
\label{fig:GradientPlots}}
\end{center}
\end{figure*}

Figure~\ref{fig:GradientPlots} shows the logarithmic colour gradients plotted against $M_V$ and $\log{\sigma_0}$. 
All elliptical galaxies have strong positive $\nabla_{FN}$ , but the gradient does not correlate strongly with either 
$M_V$ or $\log{\sigma_0}$. $\nabla_{NJ}$ is predominantly negative, and correlates strongly with $M_V$ and in 
particular with $\log{\sigma_0}$ (SRCC = 0.80; $p < 0.0001$), in the sense that the largest negative gradients
are in galaxies with $\log{\sigma_0} < 2.35$. 

Colour gradients in early-type galaxies, particularly those involving the bluer observable bands, have long been thought to be a measure of a 
metallicity gradient, and thus the dependence of gradients upon kinematic parameters has been used as a probe of galaxy formation models. 
Peletier et al. (1990) find no significant correlation between $\nabla_{UR}$ and $M_V$, although they cover a wider range in absolute magnitude
than we do and there is some indication in their Figure 12 that the gradient is strongest for galaxies with $-20 > M_B > -24$ and weaker at both brighter
magnitudes, and at fainter ones which we do not cover. La Barbera et al. (2005) find no significant dependence of $\nabla_{gr}$ upon $M_R$, however they
do find an environmental dependence, with steeper gradients away from rich clusters. La Barbera et al. (2010) define a composite colour gradient $\nabla_*$
and find very weak dependence upon $\log{\sigma_0}$ in a sample of early-type galaxies from the Sloan Digital Sky Survey (SDSS). 
Tortora et al. (2010) analyse the colour gradients in a much larger sample
of galaxies from SDSS. For the early-type galaxies in their sample they find a complex dependence 
of $\nabla_{gi}$ upon $\log{\sigma_0}$, with the strongest gradients occurring at $log{\sigma} \sim 2.0$, with weaker gradients in higher velocity dispersion 
galaxies, and weaker or even positive gradients at $\log{\sigma_0} \leq 2.0$. Tortora et al. (2010) and La Barbera et al. (2010) concur that  metallicity gradients 
are the main driver of the colour gradients, with age a secondary factor. This is borne out by a number of studies of spectroscopically determined gradients in 
stellar population parameters (Kobayashi \& Arimoto 1999; Ogando et al. 2005; Spolaor et al. 2009).

A further complication with the interpretation of our trend of $\nabla_{NJ}$ with $\log{\sigma_0}$ as evidence of a dependence on the metallicity gradient is
the possibility that the UV excess stars will contribute some flux in the NUV band as well as in FUV. Although $\nabla_{FN}$ does not correlate strongly with 
$\log{\sigma}$, the gobal value of (FUV-NUV) does, and this, combined with the overall positive gradients in this colour could lead to some leakage
of the FUV excess flux into the NUV band, preferentially at small radii and for galaxies with large $\log{\sigma_0}$. Rawle et al. (2008)
discuss the effect of the FUV excess upon the scatter in the (NUV-J) colour-magnitude relation, and show that it does not contribute
significantly to this scatter.

\subsection{Stellar Populations} 
\label{sec:stellarpops}

For all except four of our galaxies there are estimates of stellar population parameters $\log{(Age)}$, [Z/H] and [$\alpha$/Fe] in the literature.
There are differences in methodology between these sources, 
in terms of the spatial extent of the region observed, and the model spectra used to convert line index measurements into 
population parameters.  Ogando et al. (2010) use the SSP models of Thomas et al. (2003) to convert the line strength measurements
presented in Ogando et al. (2008) to population parameters. These line strengths were measured from spectra taken with a 4.1 x 2.5
arcsec slit. Annibali et al. (2007) use a new set of $\alpha$-enhanced SSP models, to fit to line strength measurements 
presented in Rampazzo et al. (2005) and Annibali et al. (2006), these line index measurements are within an aperture of 
$R_e/8$. Annibali et al. (2007) present values of Z, which we convert to [Z/H] using Z$_{\odot}$ = 0.018. Kuntschner et al. (2010)
use the SAURON integral-field spectrograph to derive line index maps, and fit the stellar population models of Schiavon (2007)
to these. They present derived stellar population parameters within $R_e/8$ and $R_e$, we use the former.
Spolaor at el. (2010) report stellar population parameters as derived in a number of
papers by the Swinburne group, for the galaxies in our sample the original sources are S\'anchez-Bl\'azquez et al. (2007; Reda at al. (2007);
Spolaor et al. (2008); Brough et al. (2007) and Proctor \& Sansom (2002).
Their values are measured within $R_e/8$, and measured line indices are converted to stellar population parameters using
the models of Thomas et al. (2003). Serra et al. (2008) present stellar population parameters within $R_e/16$ derived using
the stellar population models of Bruzual \& Charlot (2003), and the models of the index response to [E/Fe] presented by
Lee et al. (2009).  Thomas et al. (2005) measure line indices within $R_e/10$, and derive stellar population using their own
SSP models.  Loubser et al. (2009) measure line indices within a slit of 1 arcsec by $R_e/8$, and convert these to population
parameters usingthe SSP models of Thomas et al. (2003).

Given the differences in observational technique, region and SSP models, we attempt to adjust the population parameters to a 
standard scale, using galaxies in common between the samples (all of the samples are of course larger than just those galaxies
for which there are GALEX data). However even with the complete samples, the overlaps between studies are small and inconsistent.
Moreover the study with the largest sample, and the most galaxies in common with other studies (Thomas et al. 2005) is also the one
for which the rms differences with other studies are the greatest. Although we find no systematic differences between the values of 
[$\alpha$/Fe] found by different authors, the values of [Z/H] found by Kuntschner et al. (2010) are low compared with other
sources. We attempt to standardise the population parameters by 
adding 0.21 to the Kuntschner et al. (2010) values of [Z/H]. We are not making the judgement 
that the resulting values are in any way more correct, but only that they are more comparable to the other studies. 

We find that the values of Log(Age) from different sources have large systematic and random differences, and cannot derive
consistent transformations between sources. We do not consider Age any further in the current study, to do so we suggest 
would require re-observing all of our sample of galaxies with a common observational setup, and analysis of the data with a common
fitting routine and set of models.

Where [Z/H] and [$\alpha$/Fe] are given by more than one source we combine these using an {\em unweighted} mean
as the errors provided by the different sources are not derived in a sufficiently consistent way to use them for weighting.
We list the final values we have used and their sources in Table~\ref{tab:properties}.

\begin{figure*}
\begin{center}
\includegraphics[width=17.5cm]{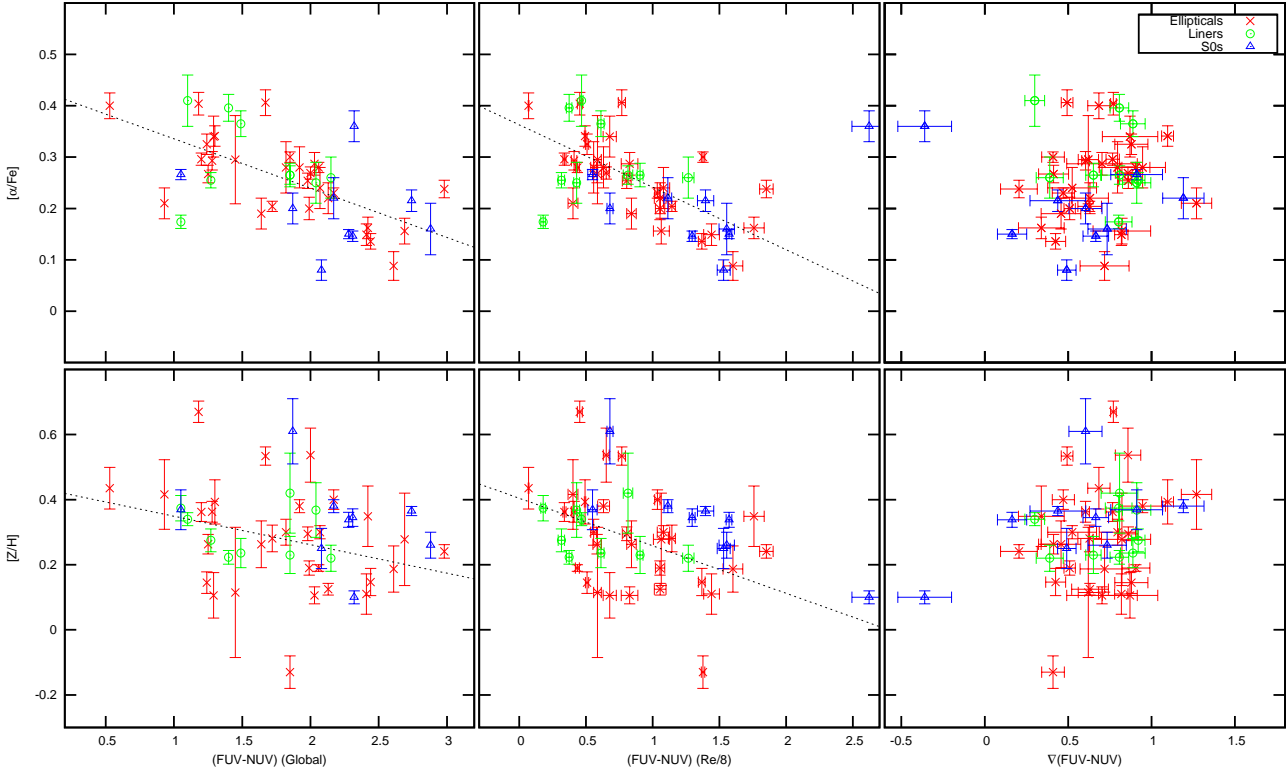}
\caption{Dependence of stellar population parameters [Z/H] and [$\alpha$/Fe] upon global and central (FUV-NUV) colour
and colour gradient. Both [Z/H] and [$\alpha$/Fe] correlate strongly with (FUV-NUV) colour, in the sense that galaxies
which are bluer in (FUV-NUV) are more metal rich and more $\alpha$-enhanced, however there is no clear correlation
between the gradients and either population parameter (right hand column).
\label{fig:PopsPlots}}
\end{center}
\end{figure*}

In Figure~\ref{fig:PopsPlots}
we plot the global colours and logarithmic colour gradients against each of the stellar population parameters. In this plot [Fe/H] in the second column of panels
is calculated from the other parameters using [Fe/H] = [Z/H] - 0.94[$\alpha$/Fe] (Thomas et al. 2003). There is a fairly well established correlation between 
(FUV-NUV) and the $Mg_2$ line index (Burstein et al. 1988; Donas et al. 2007) although Loubser \& S\'anchez-Bl\'azquez (2010) do not find such 
a correlation in their sample of Brightest Cluster Galaxies. We find significant negative correlations between both global and central (FUV-NUV) and both [Z/H] 
and [$\alpha$/Fe]. The correlations with central colour are slightly tighter (SRCC = --0.45; $p = 0.012$ for [Z/H]
SRCC = --0.66; $p = 0.0001$ for [$\alpha$/Fe]). The latter correlation suggests that 
the FUV excess may be driven by $\alpha$-element abundance rather than overall metallicity, indeed there is no significant correlation at all between either
global or central (FUV-NUV) and [Fe/H] (SRCC = --0.009; $p = 0.63$ for the central value). 

$\nabla_{FN}$ does not correlate with any stellar population parameter, but $\nabla_{NJ}$ (not shown in Figure~\ref{fig:PopsPlots}) 
correlates with [Z/H] (SRCC = 0.47; $p = 0.009$), which is
shown by a multiple regression  to be entirely a consequence of the strong dependence of both of these parameters on $\sigma_0$. 

\subsection{The S\'ersic index}
\label{sec:sersicindex}

\begin{figure*}
\begin{center}
\includegraphics[width=17.5cm]{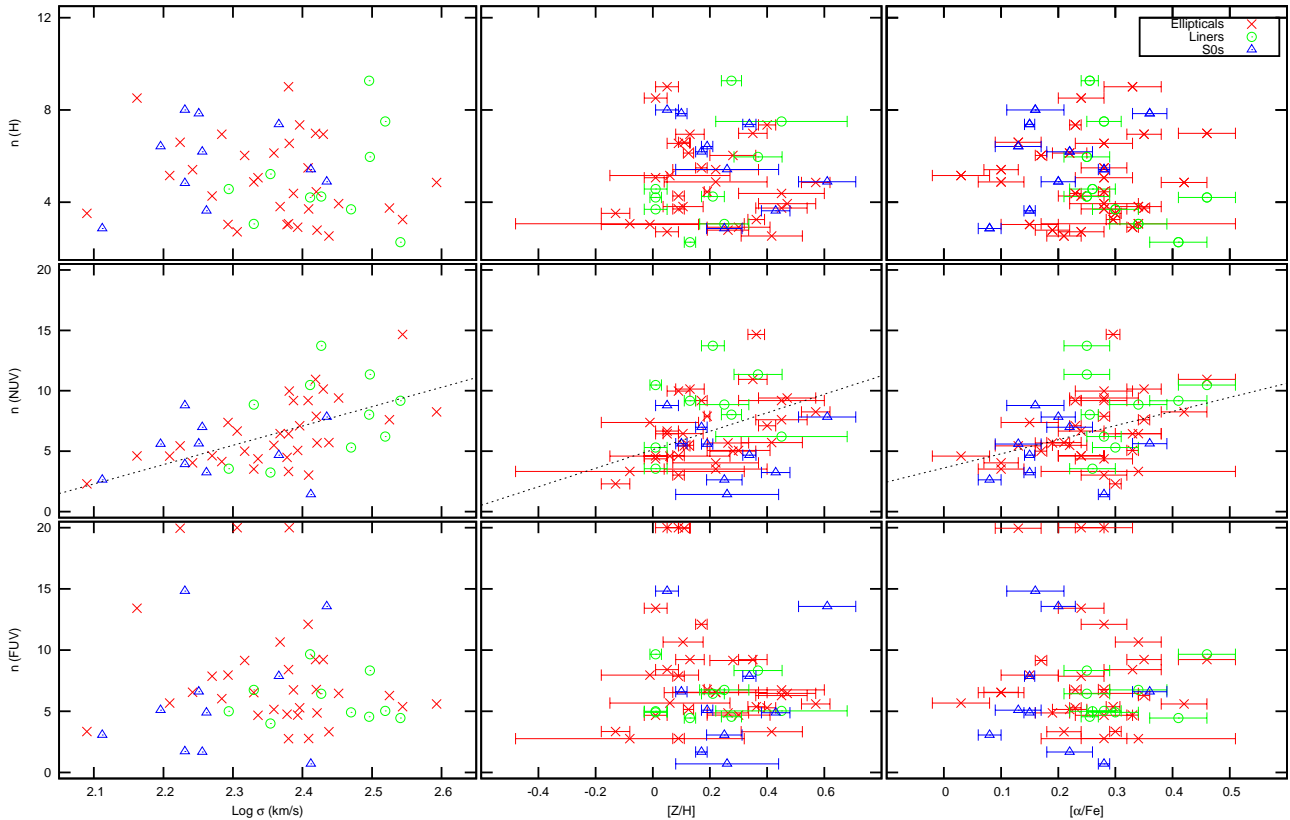}
\caption{Dependence of the S\'ersic indices from the \textsc{Galfit} fits with $n$ free, upon velocity dispersion
and stellar population parameters. The relationship between $n$ and [$\alpha$/Fe] found by
Marino et al. (2011) appears to be mostly driven by NUV passband (middle row).
\label{fig:NPlots}}
\end{center}
\end{figure*}

Marino et al. (2011) find a relationship between  [$\alpha$/Fe] and the S\'ersic index $n$, with galaxies with $n > 4$ having 
[$\alpha$/Fe] $>$ 0.15, while galaxies with smaller $n$ have a range of [$\alpha$/Fe]. They suggest that this reflects a more
fundamental relationship between [$\alpha$/Fe] and galaxy mass, which is related to the formation timescales of early-type
galaxies. The values of $n$ used by Marino et al. (2011) are an average of those determined in NUV, FUV and r bands. In Figure~\ref{fig:NPlots}
we investigate whether this relationship is present in our sample, and whether it is particuarly driven by the value of $n$ determined
in any particular band. We find no significant correlation between $n$ as measured in either FUV or H bands with either $\log{\sigma_0}$
or any stellar population parameter, but $n$ measured in the NUV band has a strong positive correlation with $\log{\sigma_0}$ (SRCC = 0.65; $p < 0.0001$)
and weaker but significant correlations with both with [Z/H] (SRCC = 0.45; $p = 0.012$), and  [$\alpha$/Fe] (SRCC = 0.46; $p = 0.010$). 
We suggest that the correlation found by Marino et al. (2011) is dominated by the NUV band, and, as they suggest, is a consequence of the correlation of all of these parameters 
with the depth of the galaxy potential well. The mechanism by which the potential affects $n$ in the NUV band is related to the strong 
UV to IR colour gradients found in galaxies of lower $\log{\sigma_0}$ (Figure~\ref{fig:GradientPlots}, upper right panel) which in turn is related
to the strong metallicity gradients in galaxies of intermediate velocity dispersion.

\subsection{IR and UV residual images}
\label{sec:residualimages}

The colour gradients show that the stars contributing to the FUV excess are more centrally concentrated than the underlying
population. We now attempt to map these regions by subtracting  from the GALEX
images a model of the underlying population as determined in the H band. We use \textsc{Galfit} to 
determine $n$ and $R_e$ for the H-band image, then we scale this model to the GALEX images, 
with the high surface brightness regions masked out. The mask was constructed 
from the FUV image. Initially a copy of the FUV image was smoothed with a circular Gaussian kernel of $\sigma$ = 3 pixels
(4.5 arcseconds). All pixels in the smoothed image below a threshold were then set to zero, and the resultant image was used
as the mask image for \textsc{Galfit}, which ignores in the fit all pixels whose value in the mask is not zero. The FUV surface
brightness of the mask threshold was set between 26.0 and 27.0 mag arcsec$^{-2}$, and was dependent upon the exposure
time of the FUV image. \textsc{Galfit} was run on both FUV and NUV images using the mask determined in FUV, and with $n$, $R_e$ 
and the ellipticity and position angle constrained to the values determined in H. The galaxy centre pixel co-ordinates were constrained 
to those found from the unmasked FUV and NUV images. So effectively we scale the H-band model to the outer regions of the FUV
and NUV images, and subtract this scaled model. 

The residual images now show the spatial distribution of the sources giving rise to the FUV
excess. In Figures B1 to B6 of Appendix B, presented in the online version of the paper only, we show, 
for a subsample of our galaxies, greyscale images showing the 
FUV images, FUV residual images, and the residual images in the NUV band obtained by the same 
process. Most FUV images show an extended FUV excess over the H-band fit at some level. 

The NUV residual images show a variety of structures, and provide some indication of the source of the FUV excess emission.
In many galaxies with extended FUV excess (e.g. NGC720, NGC1404, NGC4473, NGC4697 and many other
galaxies) the NUV residual image is negative in the core, reflecting the negative (NUV-J) and (NUV-H) colour gradients caused by
a metallicity gradient and hence enhanced line blanketing in the core. In these cases we can be confident that the FUV residual does
indeed show the distribution of the stars giving rise to the FUV excess. In a number of other galaxies (e.g. NGC1407, NGC4365, NGC4649, 
NGC4839 and NGC4889) the NUV residual image shows extended positive excess NUV emission, which we attribute to recent star formation 
in the core. This will also contribute to the FUV residual, but cannot account for it all. For instance for NGC4649, which shows the strongest
positive NUV residual, we find from the fits to the residual images described below that the magnitude of the excess is 16.26 in FUV and 17.13
in NUV. Bianchi et al. (2005) present some single stellar population and continuous star formation models (their Figure 5) and they find that (FUV-NUV) 
lies in the range -0.2 to 0.0 for models with ages between 3 and 32Myr. Thus even in the unlikely event of the NGC4649 NUV excess 
being due to such a young starburst, this can still contribute at most 50\% of the excess flux in the FUV. 

Liners, such as NGC1052, NGC4278,
NGC4486 and NGC4552, show unresolved NUV and FUV excess which we attribute to the non-thermal nuclear source, in the latter two
cases there is clearly extended FUV residual emission as well. In NGC1399, in addition to a strong extended FUV residual emission, and extended
negative NUV residual, there is an excess in both bands right in the core. This occurs in the H-band as well and is an indication that the 
S\'ersic function is an inadequate fit to the surface brightness profile, there is extra light (c.f. Kormendy et al. 2009) above the S\'ersic 
fit in all wavebands in the core of this galaxy. 

For a smaller subsample, the signal-to-noise ratio in the FUV residual images is sufficient that we can use \textsc{Galfit} to determine their structure.
Our procedure is to fit a S\'ersic function to the residual image, adopting the square root of the original FUV counts image (i.e. the GALEX archive
image multiplied by the exposure time) as a noise image. The results of the fits are given in Table~\ref{tab:UVexcess}. In this Table, column 2 gives the 
extinction corrected FUV apparent magnitude of the S\'ersic model of the residual, column 3 the effective radius R$_e$ and column 4 the S\'ersic
index $n$ of the residual image. Column 5 gives the magnitude difference between the model of the residual and the model of the galaxy image,
the latter is calculated from columns 3 and 4 of Table~\ref{tab:properties}. The magnitude difference can be as little as 0.58 magnitudes, meaning that
the central residual component can contribute up to 60\% of the FUV flux, although values of 30\% - 40\% are more typical. The upper limit to this
magnitude difference of some 3.5 magnitudes, equivalent to a residual component of 4\% of total FUV flux, represents only our inability to measure smaller 
contributions with this technique. Similarly we find that R$_e$ for the residual component is much smaller than for the galaxy image in the FUV
passband, if it were not then our procedure would not recover the residual, so we cannot rule out the existence of a more extended and more luminous
FUV component with a scale length equal to that of the old stellar component of the galaxy. The S\'ersic index, $n$ is mostly in the range 0.5 - 2.0,
with some outliers near 0 and 3. 

\begin{table}
\begin{tabular}{lcccc}\hline
Galaxy&\multicolumn{3}{c}{Residual properties}&\\
&fuv$^0$&$R_e$ (arcsec)&$n$& fuv(Resid-Total)\\ \hline
NGC720& 18.42& 4.47& 0.72& 1.68\\
NGC1399& 16.56& 4.53& 1.18& 1.44\\
NGC1404& 18.47& 4.47& 0.72& 1.96\\
NGC1549& 19.90& 4.26& 0.95& 3.49\\
NGC3379& 18.33& 5.10& 1.10& 1.54\\
NGC3608& 19.66& 1.46& 1.47& 1.49\\
NGC3923& 17.91& 4.44& 2.00& 1.26\\
NGC3962& 20.05& 4.34& 0.50& 2.19\\
NGC4365& 17.70& 4.11& 1.25& 1.28\\
NGC4406& 19.68& 4.77& 0.06& 3.36\\
NGC4473& 19.45& 3.02& 1.37& 2.00\\
NGC4621& 18.03& 1.83& 3.02& 0.95\\
NGC4649& 16.26& 6.14& 1.29& 0.91\\
NGC4697& 19.75& 4.77& 0.41& 3.06\\
NGC4874& 21.11& 5.88& 0.94& 2.90\\
NGC4889& 19.49& 4.91& 1.69& 1.57\\
NGC5044& 19.22& 8.03& 0.46& 2.42\\
NGC5982& 19.32& 1.61& 3.75& 1.07\\
NGC4261& 17.95& 4.08& 1.75& 0.87\\
NGC4278& 17.58& 2.04& 3.24& 0.83\\
NGC4374& 18.24& 6.15& 0.60& 1.90\\
NGC4552& 16.84& 2.04& 2.28& 0.58\\
IC1459& 17.99& 2.91& 1.45& 1.37\\ \hline
\end{tabular}
\caption{Properties of the S\'ersic function fits to the residual resulting from subtraction of the scaled
H-band model from the FUV image\label{tab:UVexcess}}
\end{table}

\section{Discussion}
\label{sec:discussion}

Our (FUV-NUV) gradients and the residual images show that the FUV excess stars are more centrally concentrated
than the population which contributes the bulk of the light at longer wavelengths. All normal ellipticals, all liners, and all bar one of 
the starforming sample show positive (FUV-NUV) gradients, yet there is no clear correlation between the gradients and 
the absolute value of the FUV excess. The FUV excess is centally concentrated irrespective of its absolute level. 
There is no clear indication that the FUV excess depends upon environment, for instance it is not particularly strong 
in either NGC4874 in the core of the Coma cluster, or NGC4486 in the centre of the X-ray distribution in Virgo. It can be moderately 
strong in galaxies which show evidence of recent minor mergers, for instance the shell ellipticals NGC4552 (Malin 1979), NGC3608
(Forbes \& Thomson 1992), NGC3923 (Malin \& Carter 1980) and NGC5982 (Sikkema et al. 2007), but in galaxies with evidence 
of recent major mergers such as NGC474 (Turnbull et al. 1999), NGC1316 (Schweizer 1980) and NGC2865 (Hau et al. 1999), and
the kinematically decoupled core galaxy NGC596, the FUV excess is weak. 

The UV bright stars appear to be part of an old population, formed in a short period of time early in the process of galaxy formation. 
Pipino \& Matteucci (2004) and  Pipino et al. (2006, 2008, 2010) present a number of hydrodynamic models for the 
formation of ellipticals, a key prediction of these models is that star formation ceases earlier in the outer regions of the galaxies
(``outside-in'' formation). These models are successful in predicting the observed steep metallicity gradients determined from
the optical spectra ($\nabla_{[Fe/H]} \simeq -0.3$) and, by taking into account difference in initial conditions, they can reproduce
the observed variety in $\nabla_{[\alpha/Fe]}$.

However Figure~\ref{fig:PopsPlots} 
shows that the FUV excess is more a feature of $\alpha$-enhanced than of metal-rich populations so it is likely that some other factor
is driving the steepness and ubiquity of the UV-IR colour gradients. Helium abundance is a clear candidate, it has long been suspected that
enhanced helium facilitates the formation of Extreme Horizontal Branch (EHB) stars. Norris (2004) and 
Lee et al. (2005) suggest that a large enhancement in the helium abundance (${\Delta}Y \simeq 0.15$) can explain the extended blue horizontal branch
in part of the stellar population of the globular cluster $\omega$ Centauri. Sohn et al. (2006)  investigate the UV properties
of the globular clusters associated with NGC4486. Many of these clusters are very blue in (FUV-V), but in contrast with elliptical
galaxies the bluest UV colours are found for the clusters with the weakest $Mg_2$ index. Kaviraj et al. (2007) propose that this
indicates that these clusters have a small fraction of a helium enhanced population. Globular clusters are of course metal-poor populations
compared with massive elliptical galaxies, but these observations strongly suggest that large variations in helium
content can be generated by some process or processes.  Suggestions include self-enrichment by winds from the first generation of
massive AGB stars (D'Antona \& Ventura 2007) or by winds from massive rapidly rotating stars (Decressin et al. 2007). However 
whether the helium is primordial of the product of early self-enrichment, we need a mechanism to generate the strong gradients
in the FUV excess that we observe. 

A number of authors, including  Chuzhoy \& Loeb (2004) and Peng \& Nagai (2009), 
propose that helium sedimentation in cooling gas can generate enhanced helium abundances in brightest cluster galaxies. The efficiency of this 
process is a strong function of the temperature of the cooling gas (Diffusion velocity $\propto$ T$^{1.5}$, and it is supressed by turbulence 
and by small scale magnetic fields. Nevertheless both of these studies conclude that sedimentation can lead to an enhancement of Helium
by a factor up to 1.2 in cD type galaxies, where the gas temperature is $\approx$ 10$^7$K. 
At first sight this sedimentation process is an attractive candidate for the origin of the central
gradients in the FUV excess, however there are two problems with this hypothesis, The first is that it is supressed 
at high gas concentrations, as the pressure gradient causes outward diffusion which supresses 
sedimentation. Thus the helium abundance enhancement peaks at $\sim 0.1 r_{500}$, where $r_{500}$ 
is the radius within which the mean enclosed mass density is 500 times the critical density of the universe 
(Peng \& Nagai 2009). 
A further problem is the sedimentation timescale,  which is given by:
\begin{equation}
\label{tau}
\tau=12\;{\rm Gyr}\;\left(\frac{f_{g}}{0.1}\right)\left(\frac{T}{10^8
{\rm K}}\right)^{-3/2}F_B.
\end{equation}
where $f_{g}$ is the local gas fraction; T is temperature; and $F_B$ is a suppression factor due to the 
magnetic field (Chuzhoy \& Loeb 2004).  $\tau$ is likely to be a few Gyrs, and so it is difficult to set 
up a strong gradient in the early stages of galaxy formation.

Whatever the origin of the enhancement, the persistence of the gradients to the current epoch suggests that 
this population has not been disturbed by ``dry'' mergers, which would tend to smooth out the gradient. Simulations of the 
effect of mergers on the persistence of metallicity gradients have been carried out by  Kobayashi
(2004) and Di Matteo et al. (2009). The effect of mergers on a helium gradient is a somewhat simpler case, as there is
no clear mechanism for regenerating such a gradient if it is disturbed. Di Matteo et al. (2009) consider dry mergers of galaxies
of equal mass galaxies with a range of initial metallicity slopes. Although they find that a gradient can be preserved or even enhanced
through a merger, this is only true when they consider a ``primary'' galaxy merging with a ``companion'' of much steeper slope. 
In the realistic case of a merger between two galaxies of equal initial slope, they find the remnant to have 0.6 times this initial slope. 
Repeated major mergers would thus very quickly wash out the initial gradients. If the UV colour gradients are driven by 
helium abundance, the absence of a regeneration mechanism means that they provide a stronger constraint on the importance of dry 
mergers in the formation of ellipticals than do the metallicity gradients.

The weak AGN and Liners shows very little difference in their UV properties from the main sample, particularly once the central
3 arcsec aperture is taken out. It appears that a weak AGN has little effect on the global UV properties. The S0/SA0/SAB0 galaxies 
differ from the main sample in their NUV properties, in particular in $\nabla_{NJ}$, which is strongly positive in many of this sample.
However the FUV properties of this sample follow largely the same correlations as the main sample, and it seems that the FUV
excess is unrelated to any ongoing star formation activity.

\section{Conclusions}
\label{sec:conclusions}

In agreement with previous authors we find that early-type galaxies show an excess of flux in the GALEX far ultraviolet band,
compared with that expected from a normal old stellar population. There is a considerable variation in the magnitude of this excess, 
and a complex set of dependencies upon the dynamical and stellar population parameters of the galaxy.
The excess shows a stronger correlation with the central velocity dispersion than with the absolute magnitude of the galaxy. 
It also shows a definite correlation with [$\alpha$/Fe] in the old stellar population, and a weaker correlation with
[Z/H]. 

The FUV excess is more centrally concentrated than the underlying old stellar population in all galaxies except for currently starforming 
systems and recent major merger remnants. The logarithmic (FUV-NUV) colour gradient $\nabla_{FN}$ is overwhelmingly positive,
is typically 0.6 magnitudes/dex, and shows a weak positive correlation with the central internal velocity dispersion $\sigma_0$, but it
correlates much more weakly with the strength of the FUV excess itself. It does not correlate significantly with any stellar population parameter.

Most galaxies show a negative colour gradient in (NUV-J), which can be attributed to a metallicity gradient. This gradient is strongest in
galaxies with $\log{\sigma_0} < 2.35$ although we do not sample galaxies with $\log{\sigma_0} < 2.1$. 
The relationship between the S\'ersic index $n$, and [$\alpha$/Fe] in the stellar population found by Marino et al. (2011) appears  to be driven 
largely by the NUV band, and appears to be related to the strong UV to IR colour gradients in galaxies with low $\sigma_0$, and the correlation
between $\sigma_0$ and [$\alpha$/Fe].

When we treat the FUV excess as a separate component, we find that it is well fit with a S\'ersic model profile with the S\'ersic index
$n$ in the range 0 to 3.5. This component can contribute up to 60\% of the total FUV flux from the galaxy. The most likely 
origin of the FUV excess is in a population of old, probably metal poor, stars which coexist with the optically dominant
old metal-rich population in the centres of ellipticals. Helium abundance is a plausible candidate for
the property of the stellar population which drives both the variation in the strength of the excess, and the comparatively 
uniform radial gradients in that strength. Helium sedimentation in a cooling plasma in the very early stages of galaxy formation
is a possible mechanism to establish such a gradient, although the timescale required to set up such a gradient by this process
may be unfeasibly long. The persistence of strong gradients to the present day provides
strong limits of the importance of major ``dry'' mergers in the formation of ellipticals.

\section*{Acknowledgments}

GALEX is a NASA Small Explorer. We acknowledge
support from NASA for construction, operation, and science analysis
for the GALEX mission, developed in cooperation with
the Centre National d'Etudes Spatiales of France and the Korean
Ministry of Science and Technology. This publication makes use of data products from the Two Micron All Sky Survey, which is 
a joint project of the University of Massachusetts and the Infrared Processing and Analysis Center/California Institute of 
Technology, funded by the National Aeronautics and Space Administration and the National Science Foundation. This research has 
made use of the NASA/IPAC Extragalactic Database (NED) which is operated by the Jet Propulsion Laboratory, California Institute of 
Technology, under contract with the National Aeronautics and Space Administration. DC and AMK acknowledge support from the Science
and Technology Facilities Council (STFC) under grant ST/H/002391/1. We acknowledge support for JK in this research from a Nuffield 
Foundation Science Bursary. We thank Dr. Sue Percival and Dr. Phil James for many helpful discussions and comments throughout the course
of this project, and the referee for a careful reading and suggestions which have greatly improved the paper.

\clearpage

\appendix

\section{Surface Brightness and Colour Profiles}
\label{app:gradients}

In Figures A1 to A13 we present surface brightness and colour profiles for our samples. In these Figures the
value of the radius plotted on the horizontal axis is simply the midpoint in radius of the annulus, with no account taken of the increased
number of pixels near the outer edge of the annulus, nor the increased weighting of the inner points nearer to the inner edge, which 
effects tend to cancel out. In each plot the top panel is the K-band surface brightness, the centre panel is (NUV-J) and the lower panel
(FUV-NUV). All photometry is corrected for galactic extinction as described in  Section 3.1.

\setcounter{subfigure}{1}
\setcounter{figure}{0}
 \begin{figure*}
 \begin{center} 
   \includegraphics[width=16cm,trim=0cm 0cm 0cm 2cm]{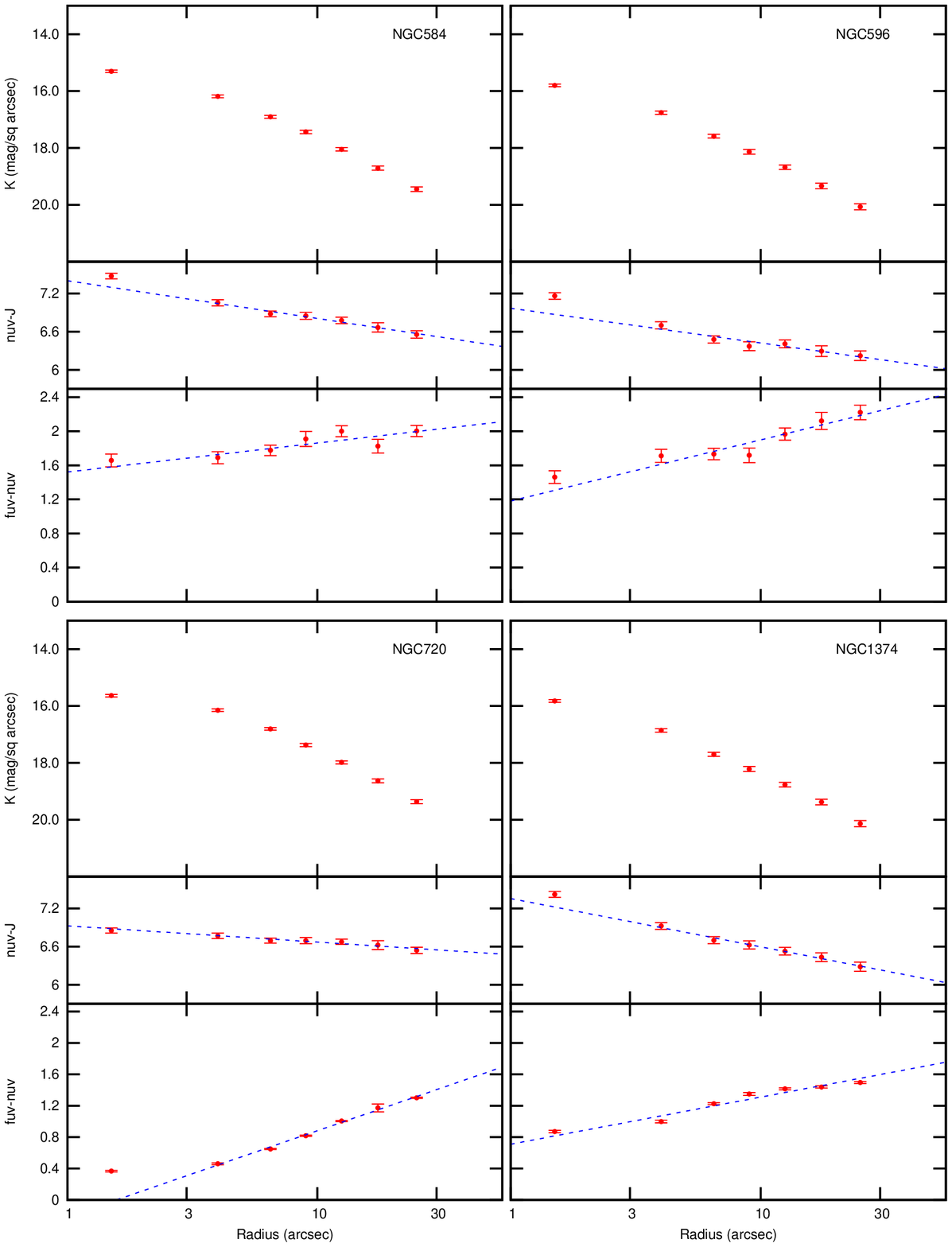} 
 \end{center} 
 \caption[]{Radial profiles in K-band surface brightness, (NUV-J) colour and (FUV-NUV) colour
for NGC584, NGC596, NGC720 and NGC1374.}
 \label{fig:fig1a}
 \end{figure*}

 \begin{figure*}
 \begin{center} 
   \includegraphics[width=16cm,trim=0cm 0cm 0cm 2cm]{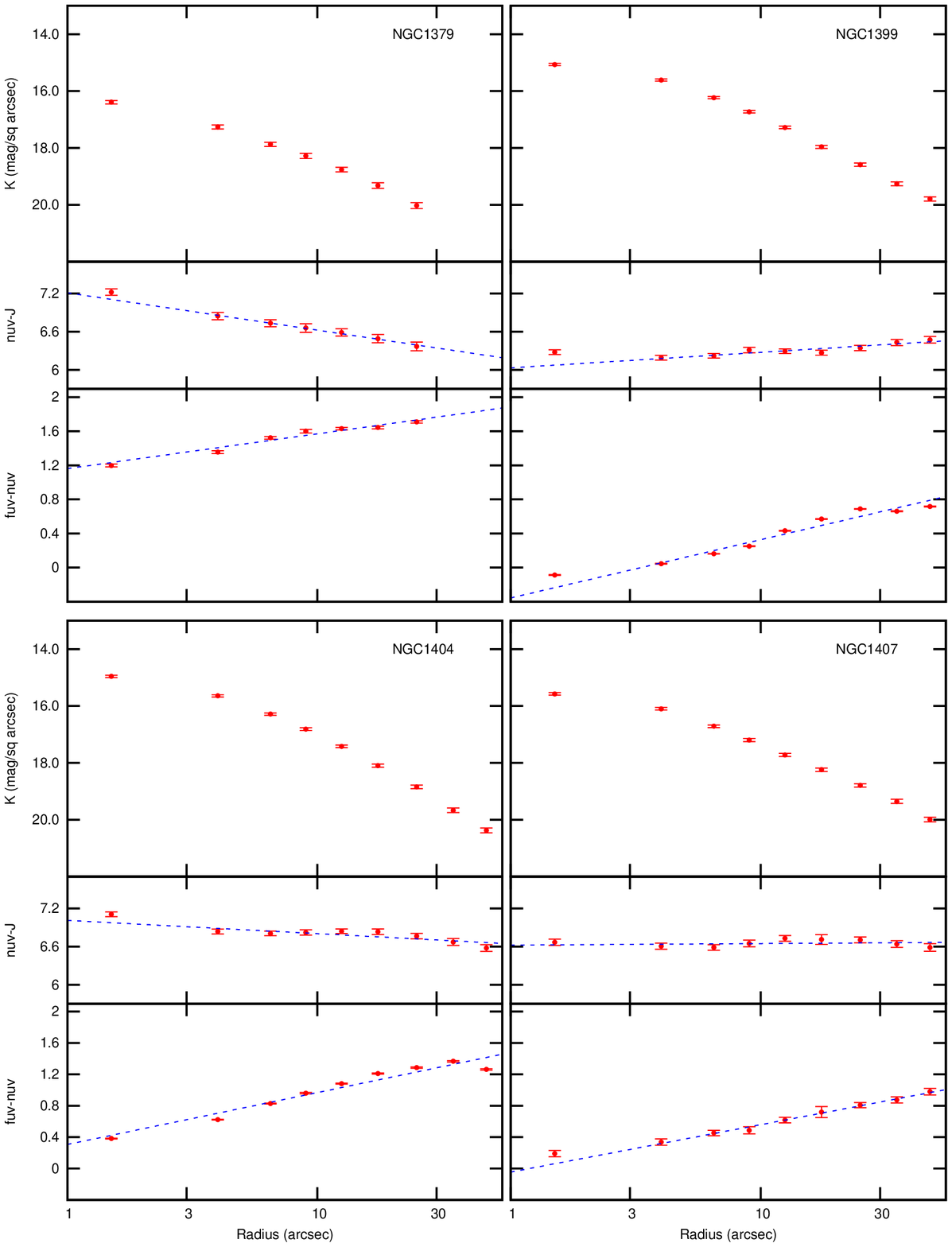} 
 \end{center} 
 \caption[]{Radial profiles in K-band surface brightness, (NUV-J) colour and (FUV-NUV) colour
for NGC1379, NGC1399, NGC1404 and NGC1407.}
 \label{fig:fig1b}
 \end{figure*}

 \begin{figure*}
 \begin{center} 
   \includegraphics[width=16cm,trim=0cm 0cm 0cm 2cm]{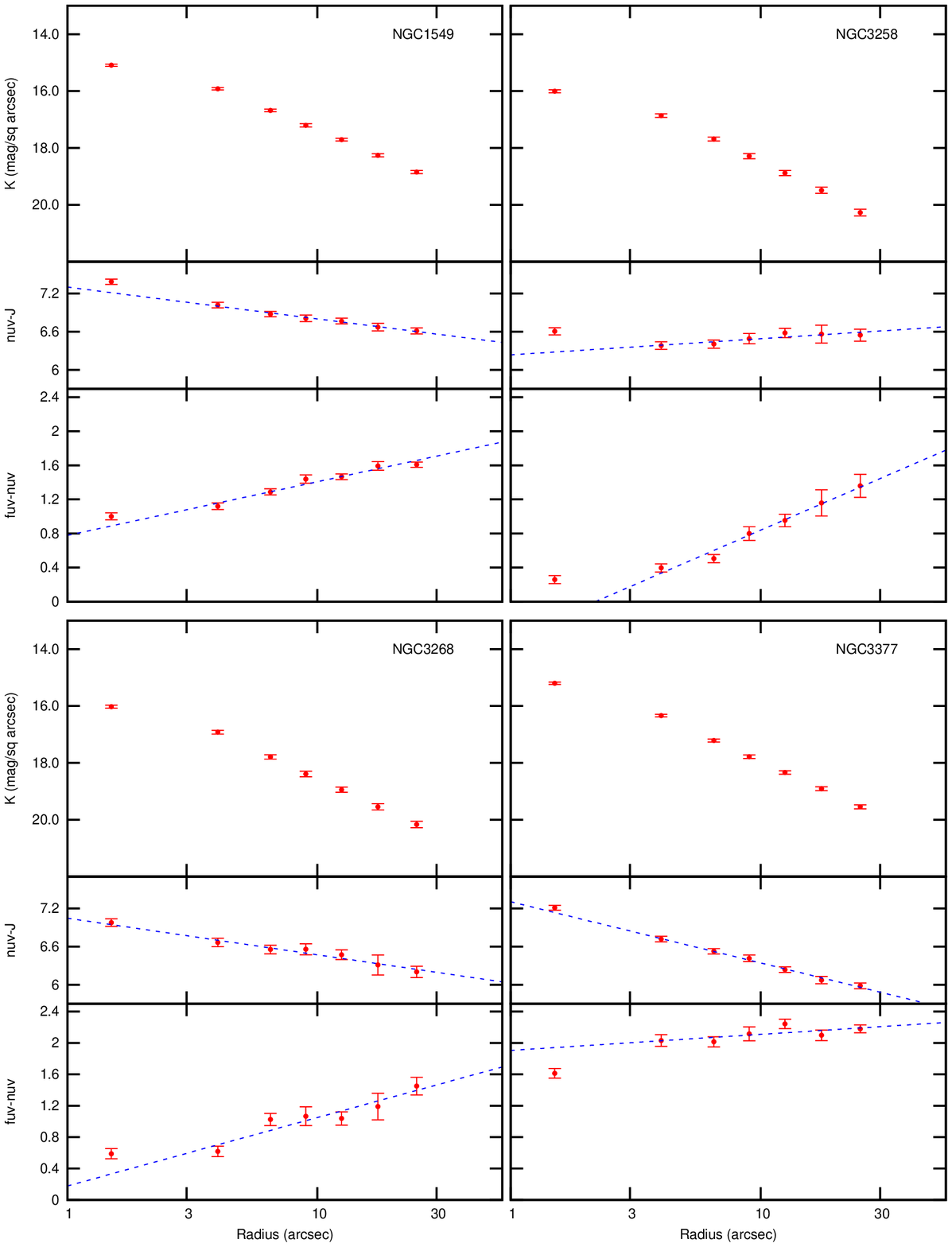} 
 \end{center} 
 \caption[]{Radial profiles in K-band surface brightness, (NUV-J) colour and (FUV-NUV) colour
for NGC1549, NGC3258, NGC3268 and NGC3377.}
 \label{fig:fig1c}
 \end{figure*}

 \begin{figure*}
 \begin{center} 
   \includegraphics[width=16cm,trim=0cm 0cm 0cm 2cm]{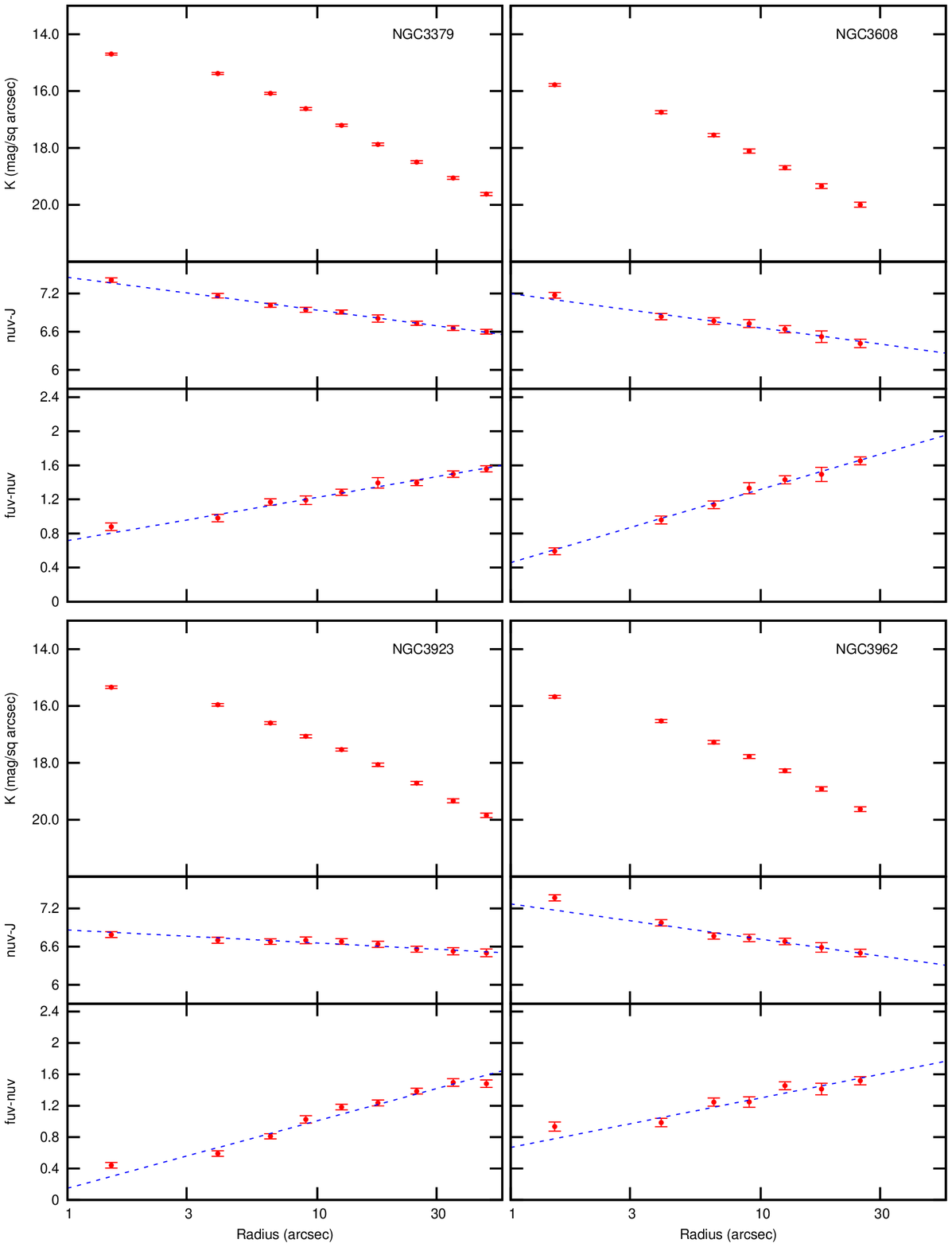} 
 \end{center} 
 \caption[]{Radial profiles in K-band surface brightness, (NUV-J) colour and (FUV-NUV) colour
for NGC3379, NGC3608, NGC3923 and NGC3962.}
 \label{fig:fig1d}
 \end{figure*}

 \begin{figure*}
 \begin{center} 
   \includegraphics[width=16cm,trim=0cm 0cm 0cm 2cm]{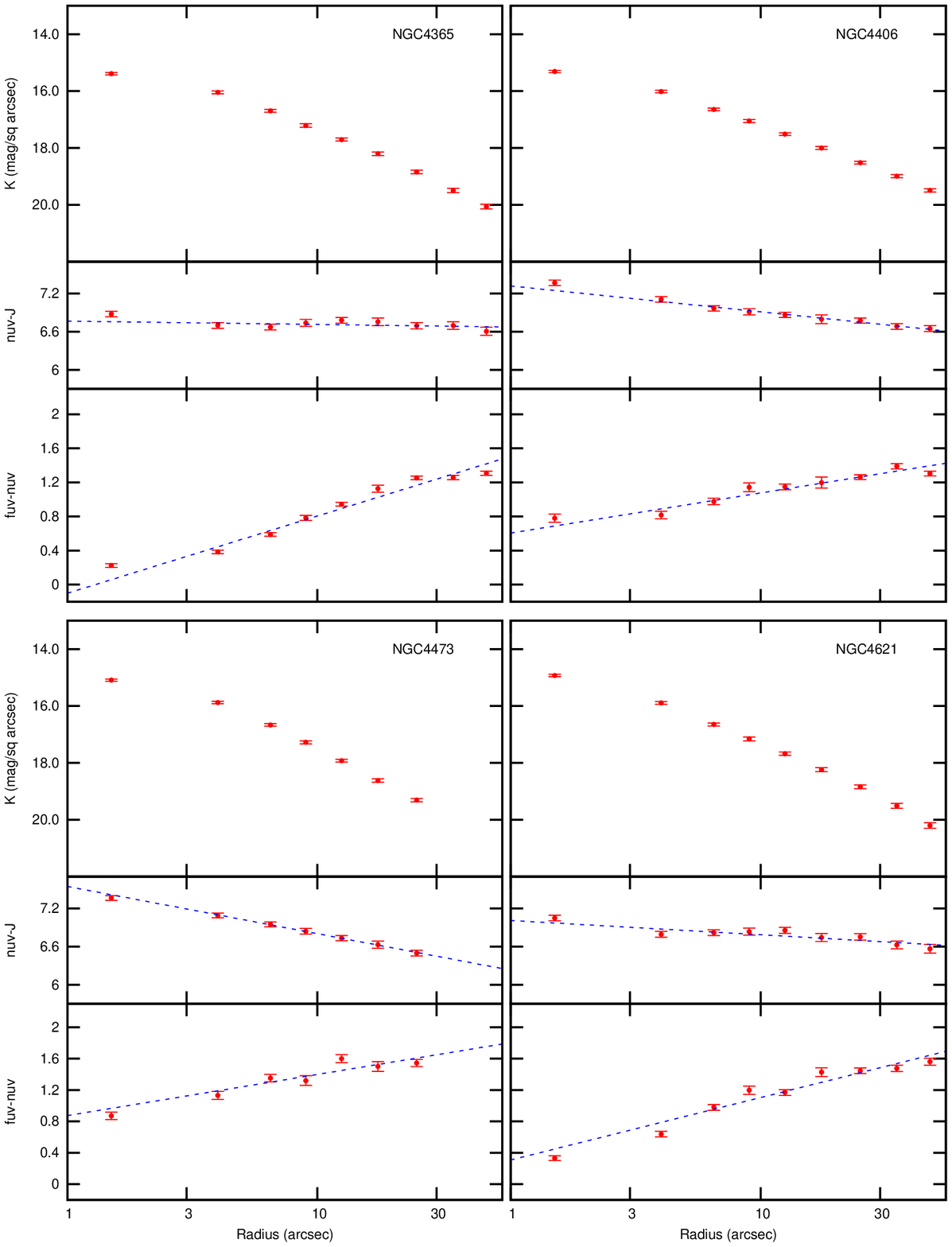} 
 \end{center} 
 \caption[]{Radial profiles in K-band surface brightness, (NUV-J) colour and (FUV-NUV) colour
for NGC4365, NGC4406, NGC4473 and NGC4621.}
 \label{fig:fig1e}
 \end{figure*} 
 
 \begin{figure*}
 \begin{center} 
   \includegraphics[width=16cm,trim=0cm 0cm 0cm 2cm]{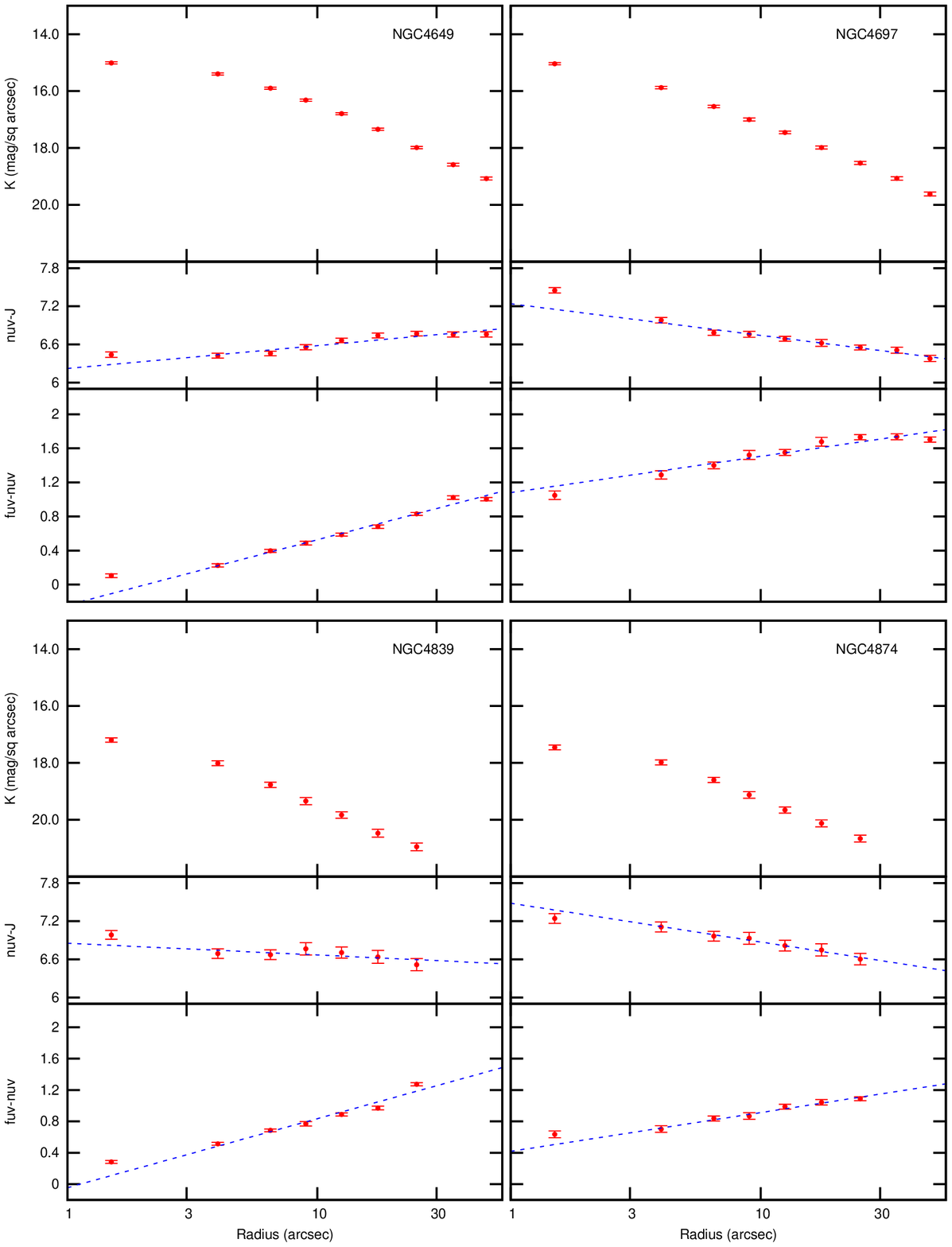} 
 \end{center} 
 \caption[]{Radial profiles in K-band surface brightness, (NUV-J) colour and (FUV-NUV) colour
for NGC4649, NGC4697, NGC4839 and NGC4874.}
 \label{fig:fig1f}
 \end{figure*}

 \begin{figure*}
 \begin{center} 
   \includegraphics[width=16cm,trim=0cm 0cm 0cm 2cm]{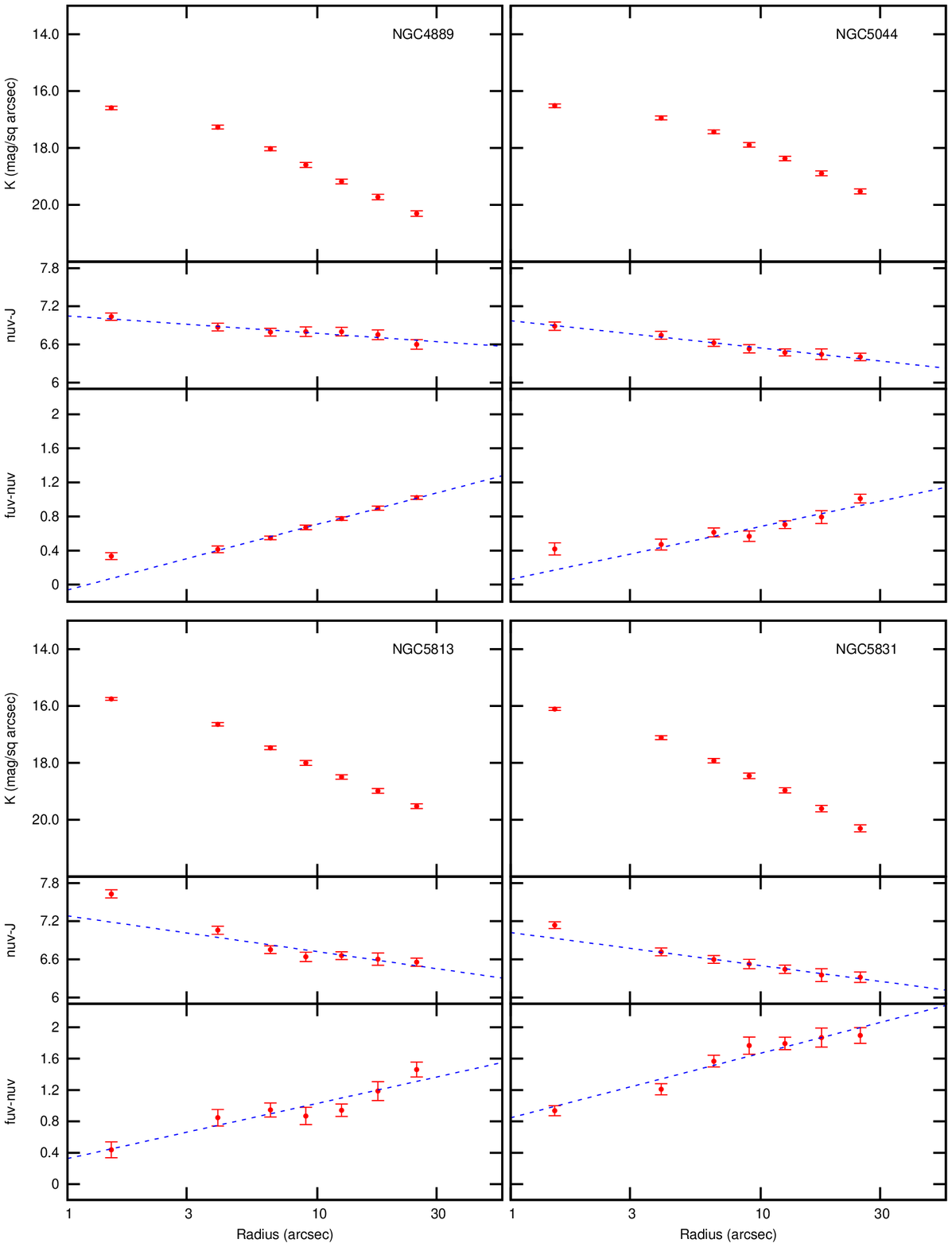} 
 \end{center} 
 \caption[]{Radial profiles in K-band surface brightness, (NUV-J) colour and (FUV-NUV) colour
for NGC4889, NGC5044, NGC5813 and NGC5831.}
 \label{fig:fig1g}
 \end{figure*} 

 \begin{figure*}
 \begin{center} 
   \includegraphics[width=16cm,trim=0cm 0cm 0cm 2cm]{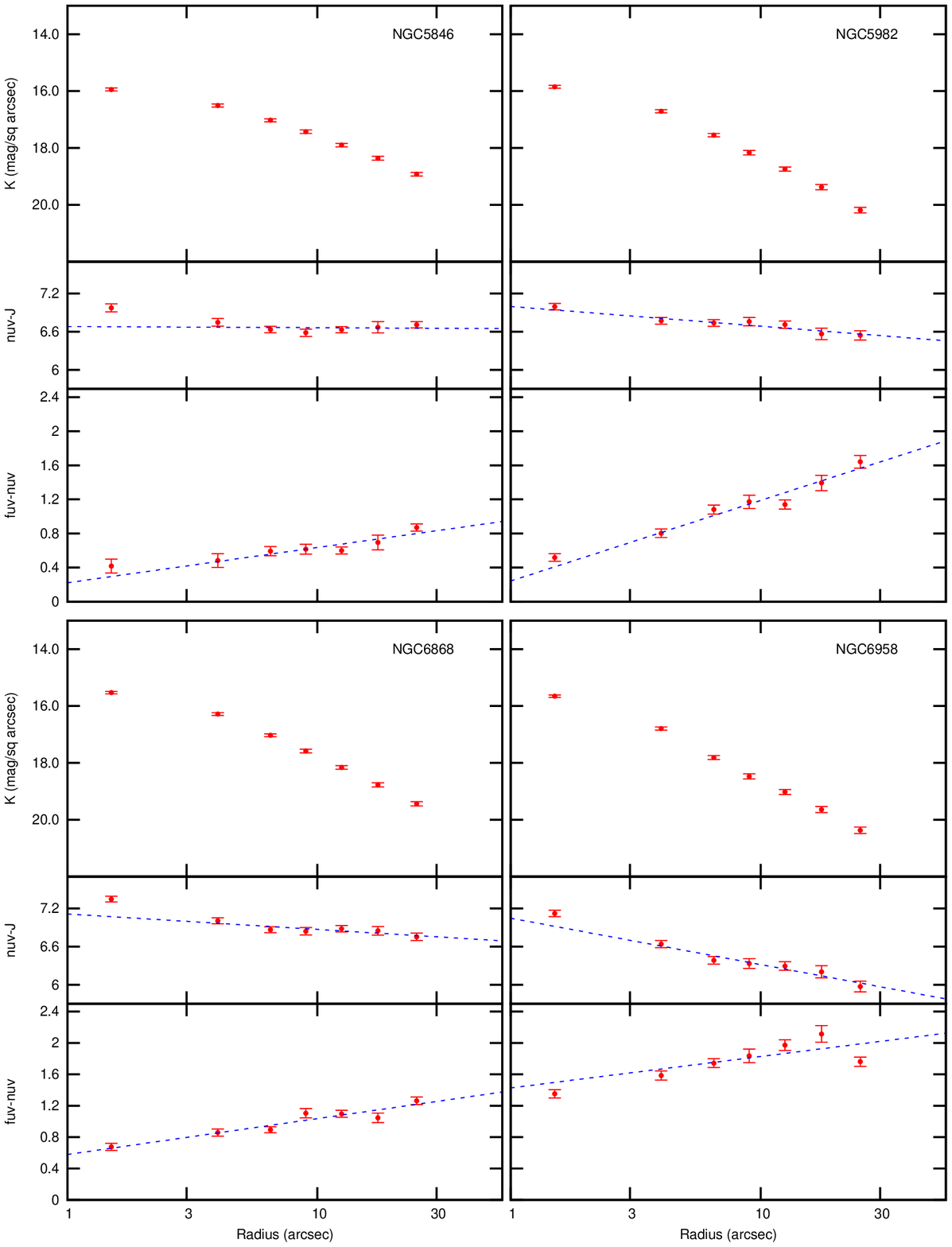} 
 \end{center} 
 \caption[]{Radial profiles in K-band surface brightness, (NUV-J) colour and (FUV-NUV) colour
for NGC5846, NGC5982, NGC6868 and NGC6958.}
 \label{fig:fig1h}
 \end{figure*} 

 \begin{figure*}
 \begin{center} 
   \includegraphics[width=16cm,trim=0cm 0cm 0cm 2cm]{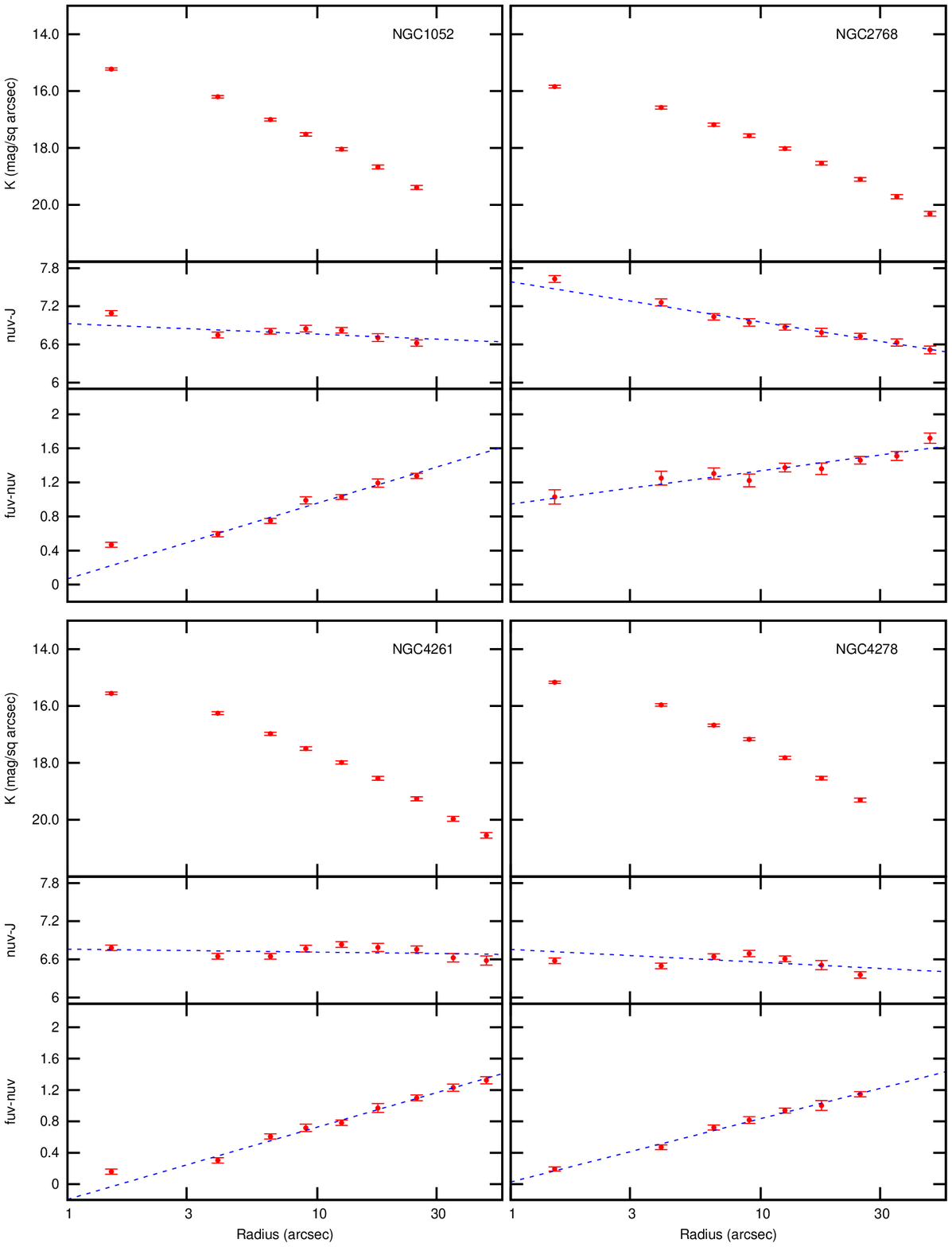} 
 \end{center} 
 \caption[]{Radial profiles in K-band surface brightness, (NUV-J) colour and (FUV-NUV) colour
for NGC1052, NGC2768, NGC4261 and NGC4278.}
 \label{fig:fig1i}
 \end{figure*} 

 \begin{figure*}
 \begin{center} 
   \includegraphics[width=16cm,trim=0cm 0cm 0cm 2cm]{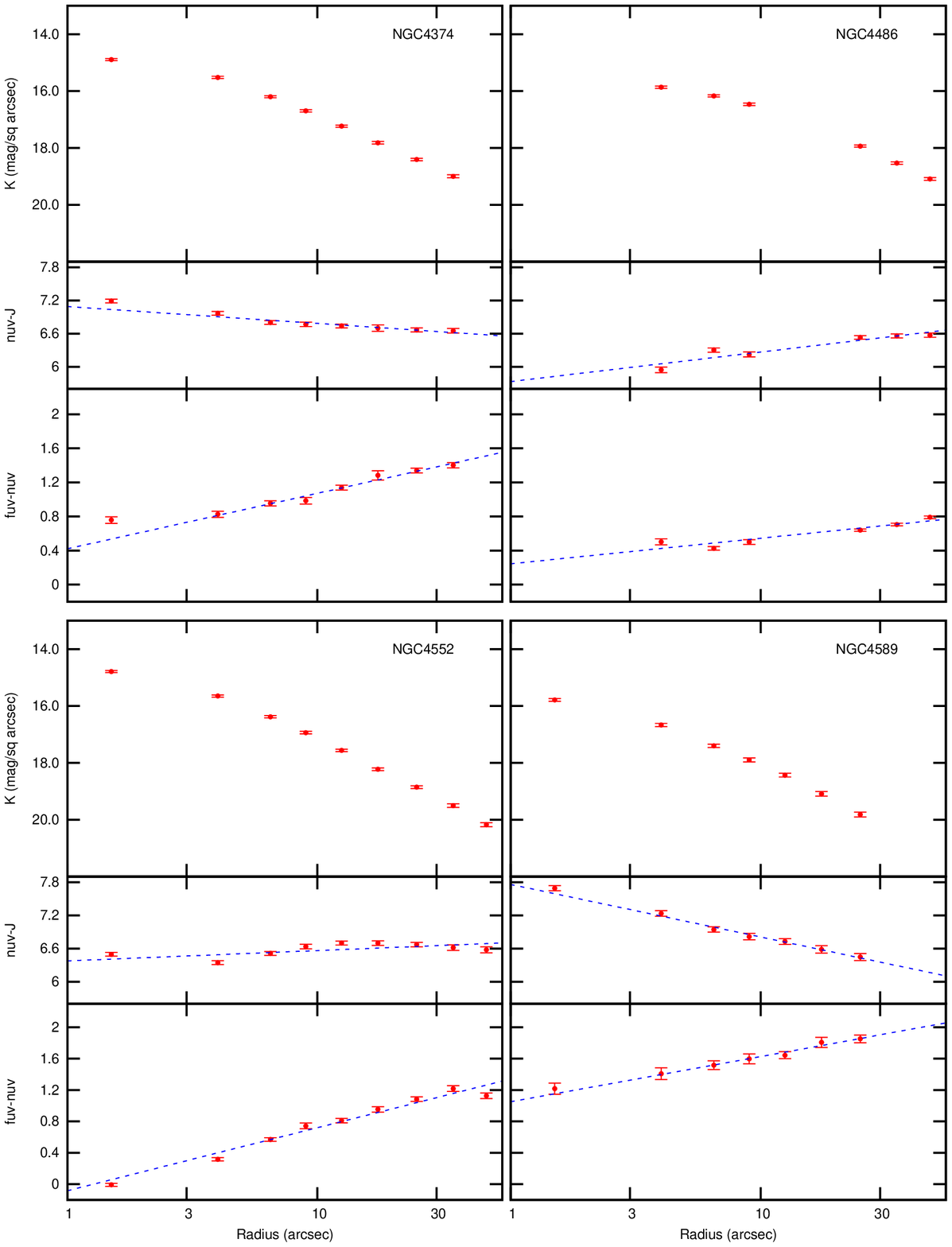} 
 \end{center} 
 \caption[]{Radial profiles in K-band surface brightness, (NUV-J) colour and (FUV-NUV) colour
for NGC4374, NGC4486, NGC4552 and NGC4589.}
 \label{fig:fig1j}
 \end{figure*} 

\begin{figure*}
 \begin{center} 
   \includegraphics[width=16cm,trim=0cm 0cm 0cm 2cm]{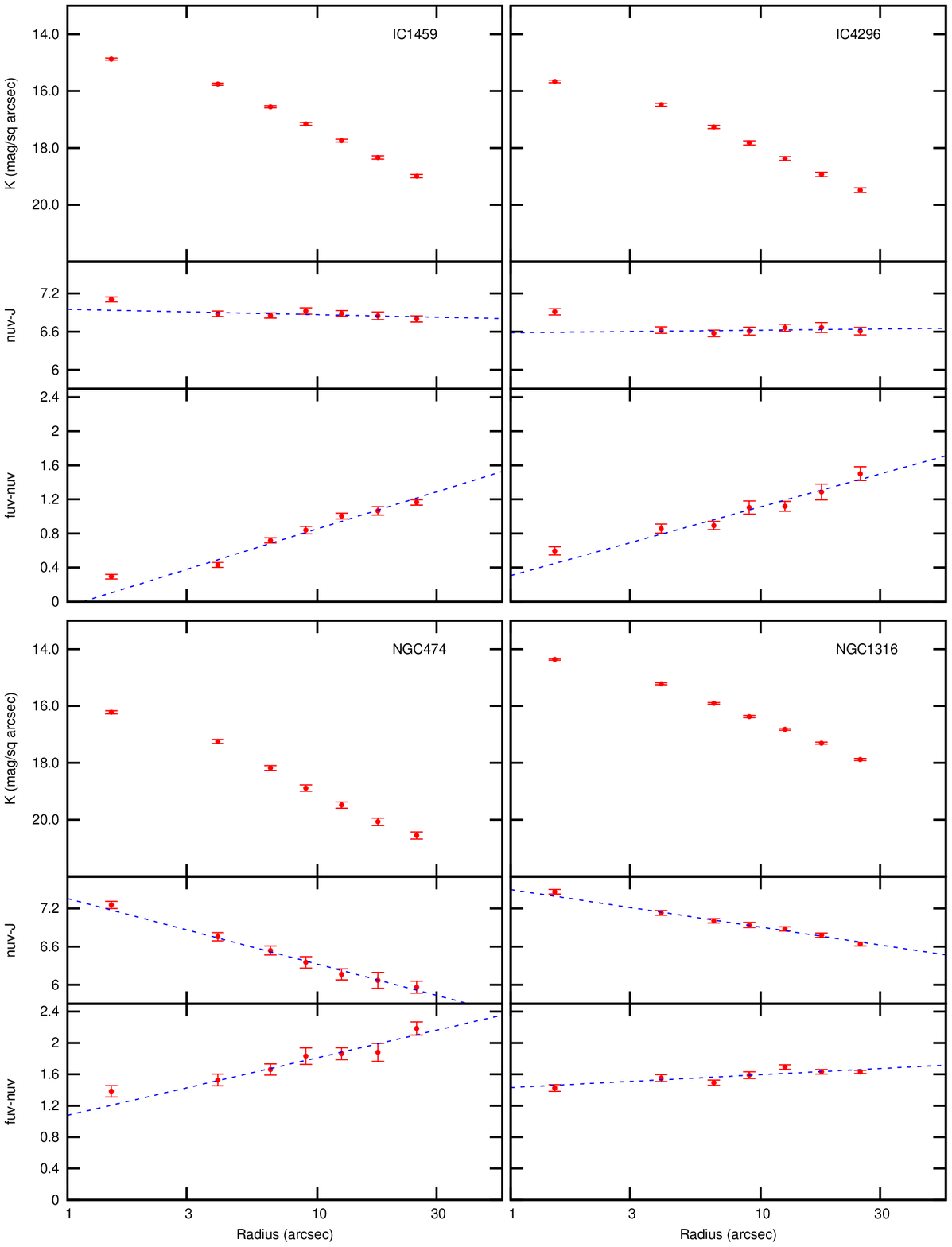} 
 \end{center} 
 \caption[]{Radial profiles in K-band surface brightness, (NUV-J) colour and (FUV-NUV) colour
for IC1459, IC4296, NGC474 and NGC1316.}
 \label{fig:fig1k}
 \end{figure*}

 \begin{figure*}
 \begin{center} 
   \includegraphics[width=16cm,trim=0cm 0cm 0cm 2cm]{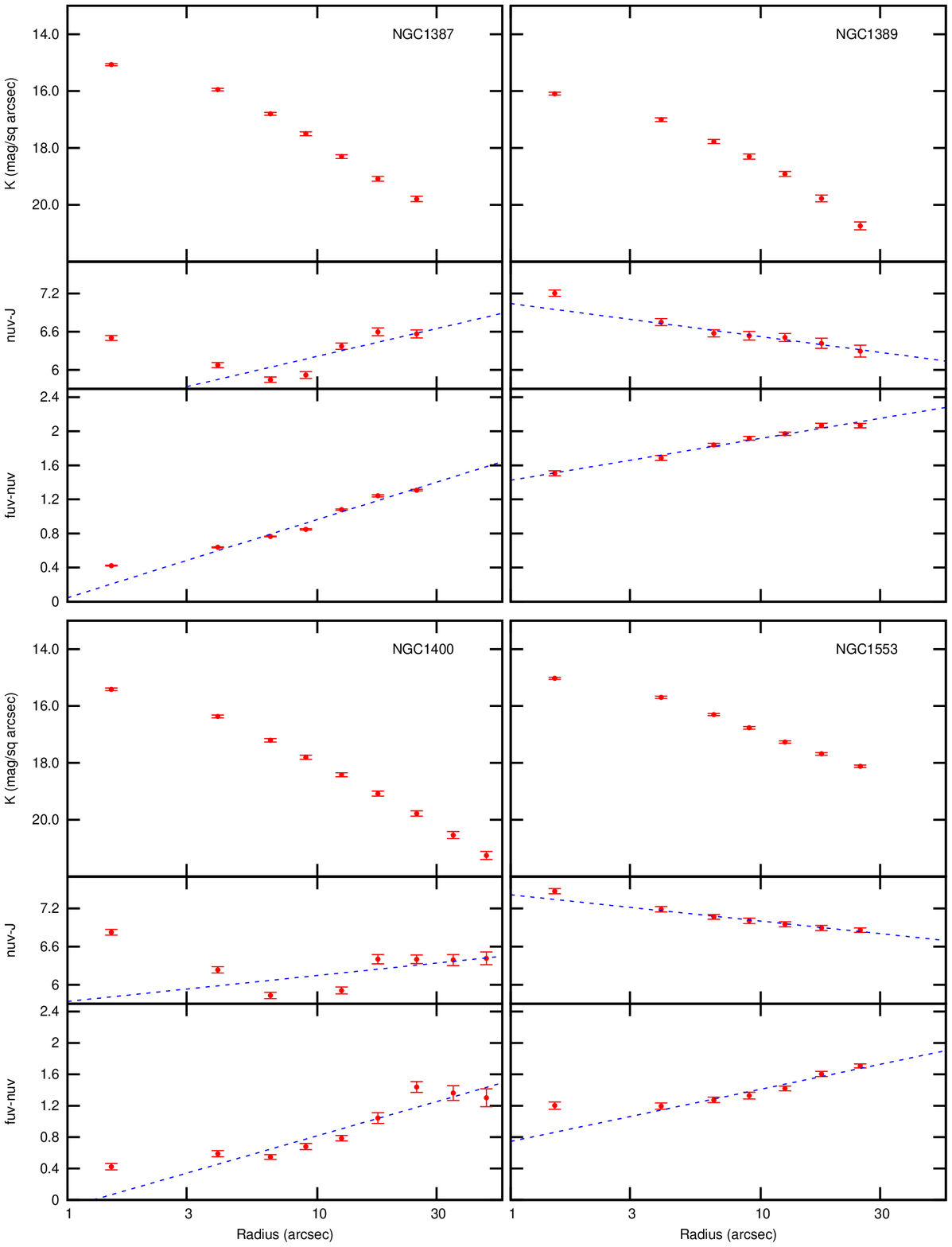} 
 \end{center} 
 \caption[]{Radial profiles in K-band surface brightness, (NUV-J) colour and (FUV-NUV) colour
for NGC1387, NGC1389, NGC1400 and NGC1553.}
 \label{fig:fig1l}
 \end{figure*}

 \begin{figure*}
 \begin{center} 
   \includegraphics[width=16cm,trim=0cm 0cm 0cm 2cm]{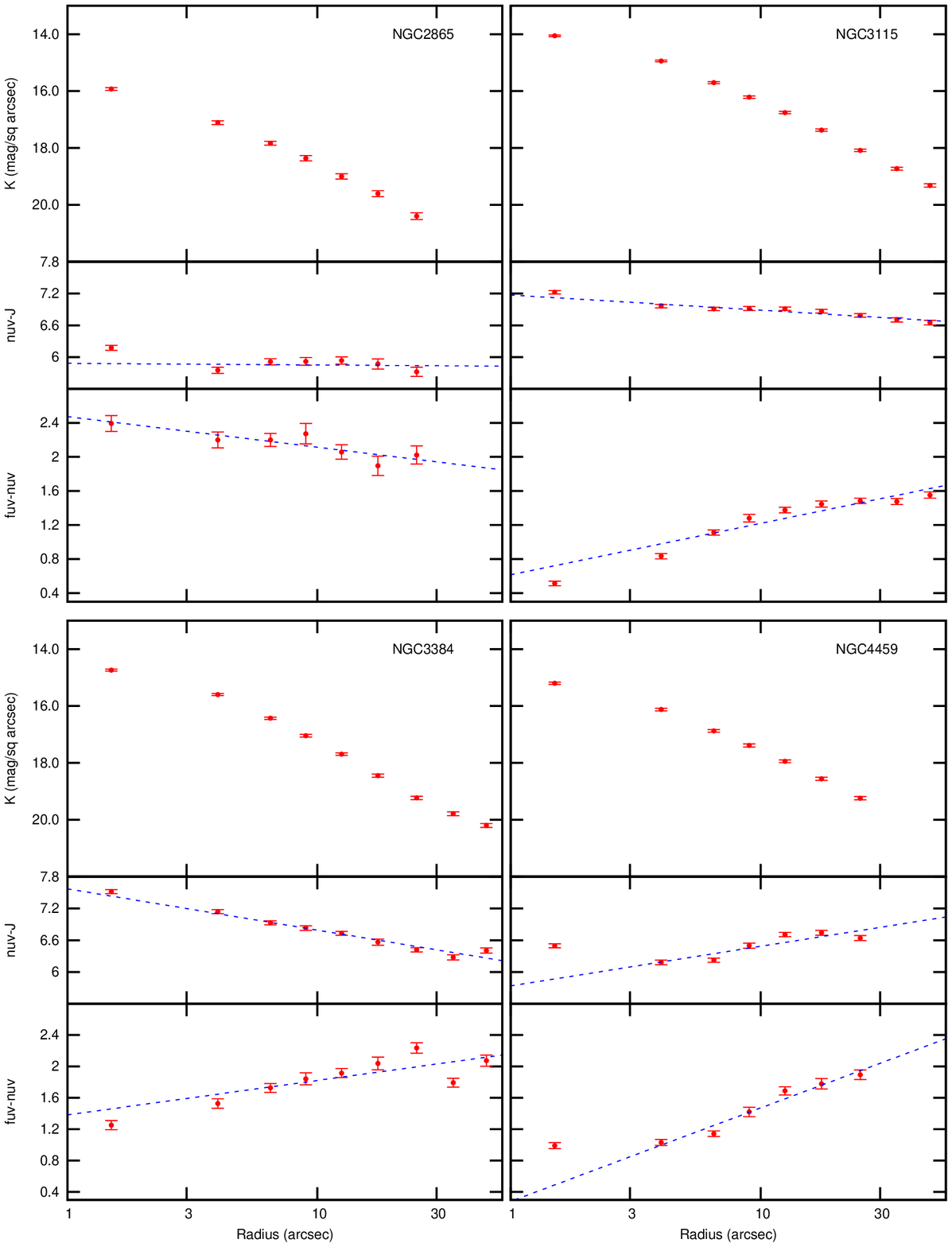} 
 \end{center} 
 \caption[]{Radial profiles in K-band surface brightness, (NUV-J) colour and (FUV-NUV) colour
for NGC2865, NGC3115, NGC3384 and NGC4459.}
 \label{fig:fig1m}
 \end{figure*}
\clearpage

\section{Residual Images}

In Figures B1 to B6 we present the residual images constructed as described in Section 3.6.
For each galaxy the left image is the NUV image as obtained from the GALEX archive, the centre panel is the residual after subtraction of the
scaled H-band model, and the right panel the NUV image after subtraction of the scaled model. Here we provide comments on the residual 
maps of individual galaxies.\\

\noindent{\em NGC720} - The NUV residual is negative, and id dominated by the metallicity gradient. The FUV residual shows an extended and
elongated UV excess distribution, and the effect of the metallicity gradient is also visible as a faint negative region around this.\\

\noindent{\em NGC1399} - The positive residual in the very centre of this galaxy is visible in all passbands, and is due to a poor fit of the S\'ersic
function to the surface brightness profile. The fit is also complicated by three other sources close to the core. There is a more extended
excess in teh FUV residual map, and a region of negative residuals due to the metallicity gradient.\\

\noindent{\em NGC1404} - The NUV residual map is dominated by the negative region caused by the metallicity gradient, and a source to the west
of the nucleus. The FUV map shows a somewhat more compact FUV excess region.\\

\noindent{\em NGC1407} - This and subsequent galaxies have much shorter exposures and hence worse signal-to-noise in the residual maps
 than the first three. The NGC1407 residual maps show an excess at small radii, in this case the residual occurs in all wavebands and is due to excess
light over the S\'ersic fit in the core. The FUV excess in this galaxy is however apparant in the colour gradients.\\

\noindent{\em NGC1549} - This galaxy shows a moderate extended FUV excess, and in the NUV band a negative metallicity gradient residual.\\

\noindent{\em NGC3379} - This also shows an extended FUV excess and a negative NUV residual.\\

\noindent{\em NGC3608} - In addition to the FUV excess, this galaxy shows a point residual in the nucleus in longer bands, it is unclear
whether this is a stellar ``extra light'' component or a weak nuclear source.\\

\noindent{\em NGC3923} - In this case there is a stronger FUV excess, and a weak positive residual in the nucleus at longer wavebands.\\

\noindent{\em NGC3962} - This galaxy has an extended FUV excess region, and a negative NUV residual in the centre.\\

\noindent{\em NGC 4365} - This galaxy has an extended FUV excess region. The positive residual in NUV is not seen at longer wavelengths,
and we attribute this to a low level of star formation in the core, rather than failure of the S\'ersic fit.\\  
 
\noindent{\em NGC 4406} - There is an extended FUV excess region, and in the NUV image a low surface brightness dwarf elliptical companion (NGC4406B)
to the NE.\\

\noindent{\em NGC4473} - This galaxy has an extended FUV excess region, and a negative NUV residual in the centre.\\

\noindent{\em NGC4621} - There is an extended FUV excess region, and a point residual caused by a poor S\'ersic fit in the core.\\

\noindent{\em NGC4649} - The fit to this galaxy is complicated by the close companion (NGC4647). However the extended NUV residual does not
appear to be caused by a poor S\'ersic fit in the H-band, but by a low level of ongoing star formation, as suggested by Magris \& Bruzual (1993).\\

\noindent{\em NGC4697} - This galaxy has an extended FUV excess region, and a negative NUV residual in the centre.\\

\noindent{\em NGC4839} - This galaxy has cD morphology, yet lies in the outer regions of the Coma cluster. It has a clear extended FUV excess region,
and a positive NUV residual which does not appear to be due to a poor S\'ersic fit in the core. We suggest that this galaxy, like NGC4649, has a low
level of ongoing star formation in the core. \\

\noindent{\em NGC4874} - This galaxy is in the centre of the Coma cluster and has cD morphology, yet it appears to have only a weak FUV excess.
The there is little residual in NUV.\\

\noindent{\em NGC4889} - This is the brightest galaxy in the Coma cluster, albeit it has elliptical rather than cD morphology. The FUV excess is much stronger
than in NGC4874. The positive residual in NUV is not caused by a poor S\'ersic fit in the core, and again we suggest that there may be a low level of star formation
in this galaxy.\\

\noindent{\em NGC5044} - The FUV excess is weak, and there are no clear residuals in the NUV image.\\

\noindent{\em NGC5846} - There is a compact companion (NGC5846A) to the south. Residual maps are similar in FUV and NUV, suggesting
that there is ongoing star formation.\\

\noindent{\em NGC5982} - The FUV excess region is compact, and the NUV residual map shows no structure. There is a faint companion 
(SDSS J153839.55+592134.3) to the north.\\

\noindent{\em NGC6868} - The FUV excess is weak and the positive residual seen in the NUV residual image occurs in all wavebands and is 
due to a poor fit of the S\'ersic function in the core, and will contribute to some extent to the FUV residual image.\\

\noindent{\em NGC1052} - There is an extended FUV excess residual, and in the NUV the Liner in the  nucleus shows as a point source.\\

\noindent{\em NGC4261} - This shows a strong extended FUV excess residual, and negative residuals in NUV.\\

\noindent{\em NGC4278} - As with NGC1052 this shows and extended FUV excess residual, and a nuclear NUV residual from the Liner. \\

\noindent{\em NGC4374} - This galaxy has a strong and extended UV excess, and negative residuals due to metallicity in NUV. There is no 
indication of any nuclear source.\\

\noindent{\em NGC4486} - M87 shows the well known non-thermal nuclear source and jet in both FUV and NUV. The extended FUV excess
region is also clearly visible.\\

\noindent{\em NGC4552} - In this case it is unclear whether the NUV residual is due to a poor fit of the S\'ersic function, or a nuclear source. 
The extended FUV excess emission is, however, clear.\\

\noindent{\em IC1459} -The residual in the NUV image occurs also at longer wavelengths, and indicates a poor S\'ersic fit. The FUV excess  
is, however, stronger and is also extended. \\

\noindent{\em IC4296} -The H-band residual in this galaxy is positive in the very centre, however the observed NUV residual is extended, and 
is more likely to be due to ongoing star formation. The FUV residual map is very similar to NUV, and the FUV excess is weak, although the 
colour gradient shows that it is present.\\

\setcounter{figure}{0}
\setcounter{subfigure}{1}

\begin{figure*}
 \begin{center} 
   \includegraphics[width=13.5cm]{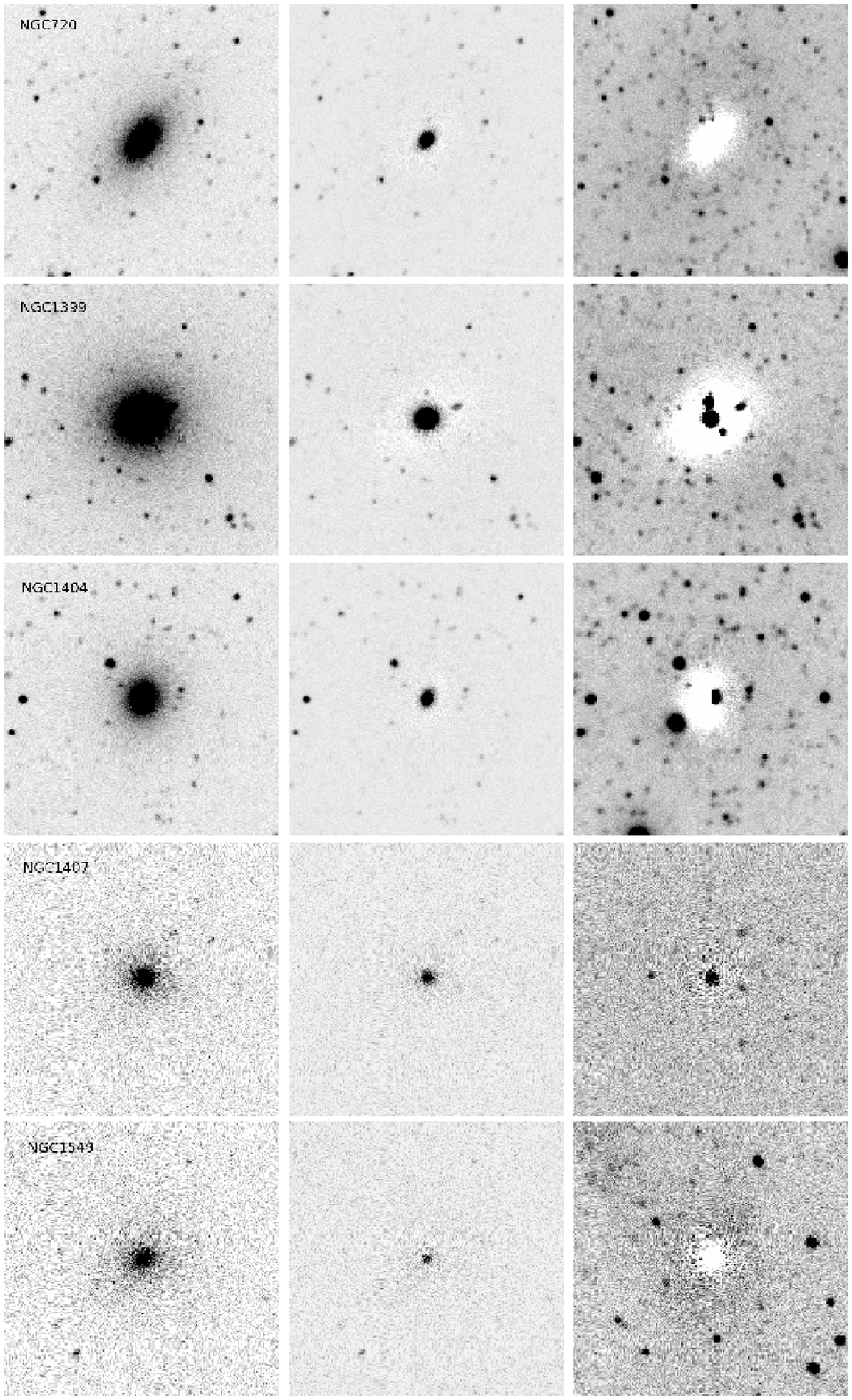} 
 \end{center} 
 \caption[]{Images of (from top) NGC720, NGC1399, NGC1404, NGC1407 and NGC1549. 
Left column: FUV image; Centre panel: FUV residual images after subtraction of scaled model determined from H-band image;
Right column: NUV residual image after subtraction of scaled model determined from H-band image. In each image north is at the top and east at the
left, and the images are each 5 arcminutes square.}
 \label{fig:fig2a}
 \end{figure*}

\begin{figure*}
 \begin{center} 
   \includegraphics[width=13.5cm]{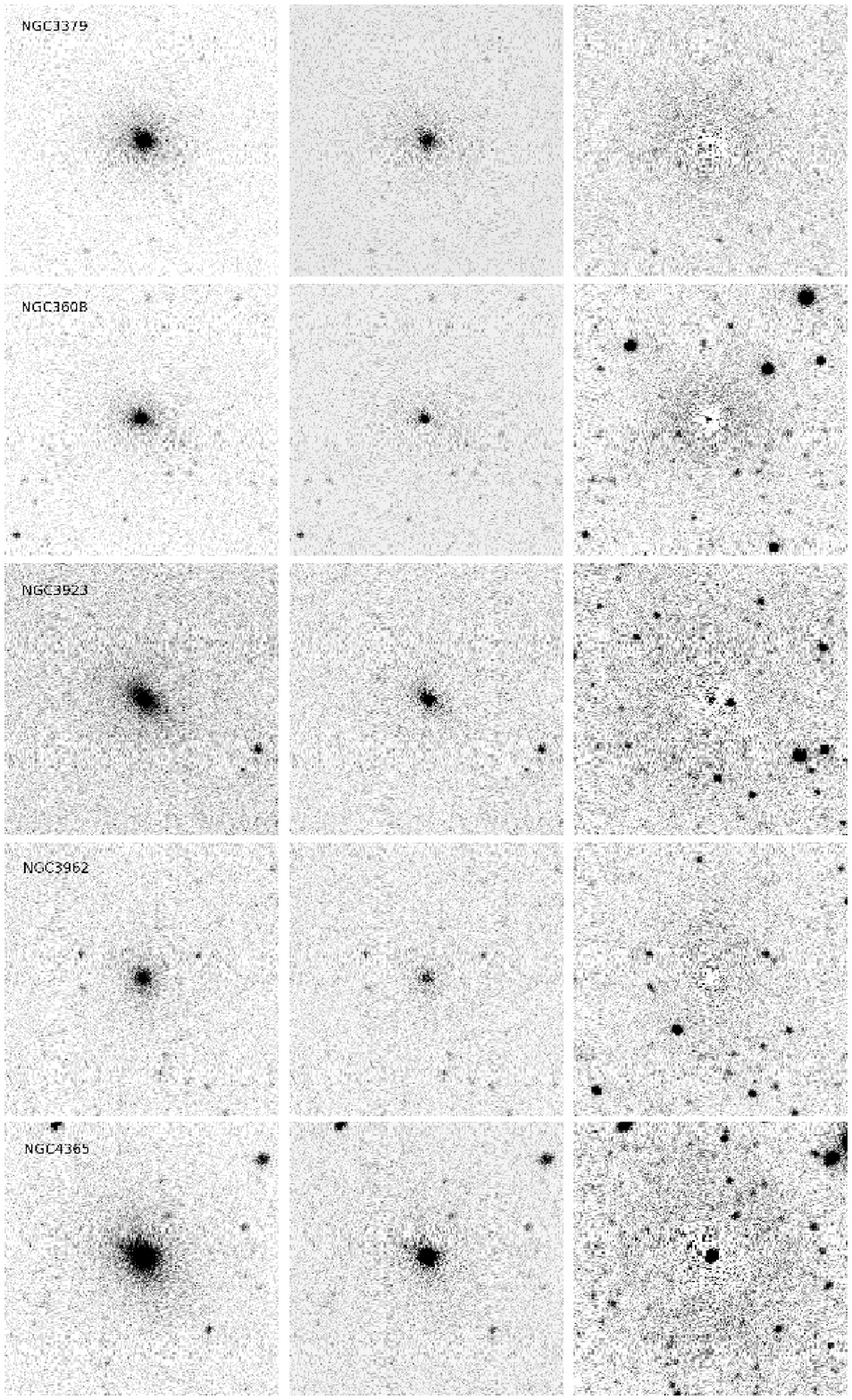} 
 \end{center} 
 \caption[]{FUV, FUV residual and NUV residual images of (from top) NGC3379, NGC3608, NGC3923, NGC3962 and NGC4365. Columns as in Figure B1}
 \label{fig:fig2b}
 \end{figure*}

\begin{figure*}
 \begin{center} 
   \includegraphics[width=13.5cm]{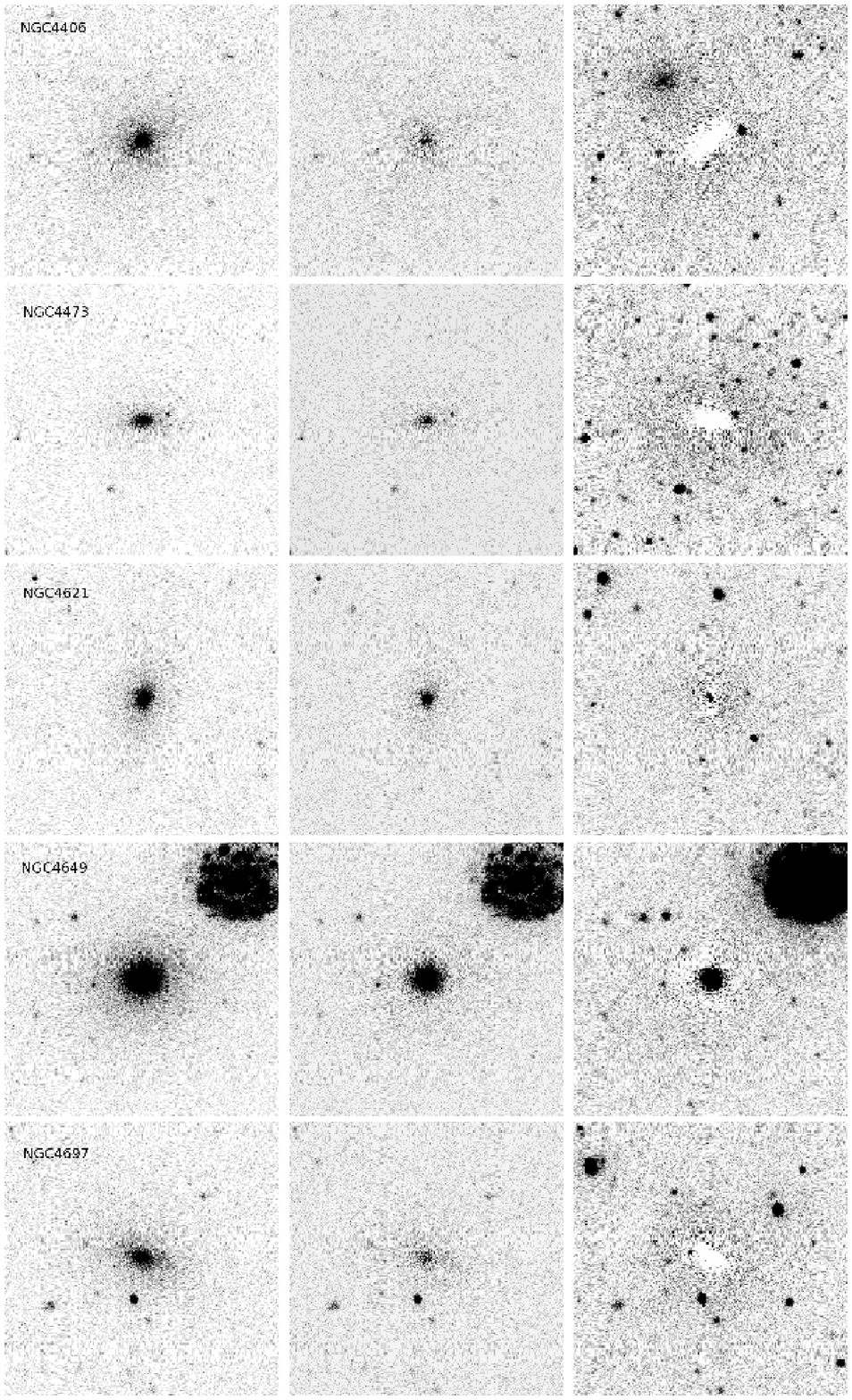} 
 \end{center} 
 \caption[]{FUV, FUV residual and NUV residual images of (from top) NGC4406, NGC4473, NGC4621, NGC4649 and NGC4697. Columns as in Figure B1}
 \label{fig:fig2c}
 \end{figure*}

\begin{figure*}
 \begin{center} 
   \includegraphics[width=13.5cm]{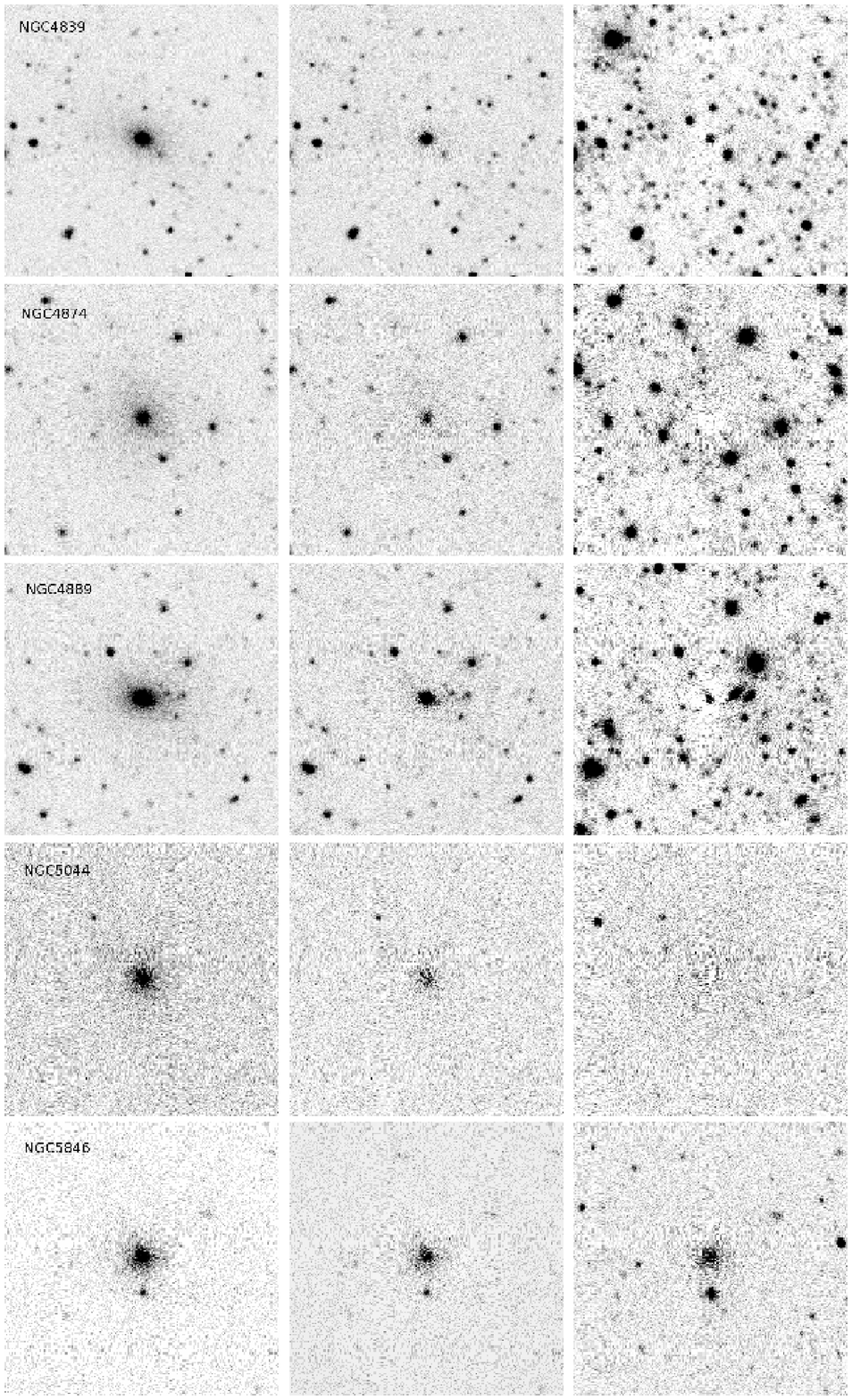} 
 \end{center} 
 \caption[]{FUV, FUV residual and NUV residual images of (from top) NGC4839, NGC4874, NGC4889, NGC5044 and NGC5846. Columns as in Figure B1}
 \label{fig:fig2d}
 \end{figure*}

\begin{figure*}
 \begin{center} 
   \includegraphics[width=13.5cm]{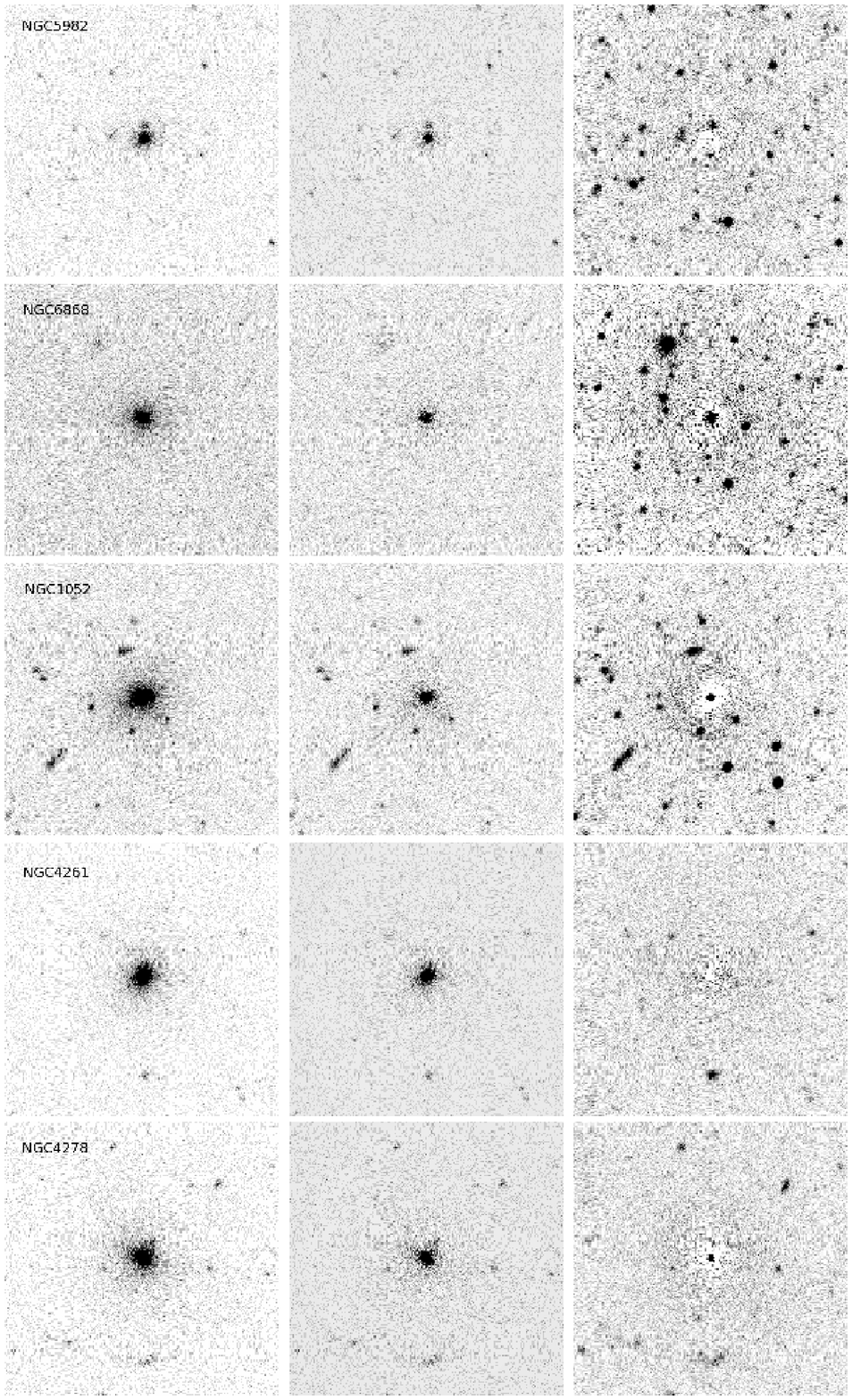} 
 \end{center} 
 \caption[]{FUV, FUV residual and NUV residual images of (from top) NGC5982, NGC6868, NGC1052, NGC4261 and NGC4278. Columns as in Figure B1}
 \label{fig:fig2e}
 \end{figure*}

\begin{figure*}
 \begin{center} 
   \includegraphics[width=13.5cm]{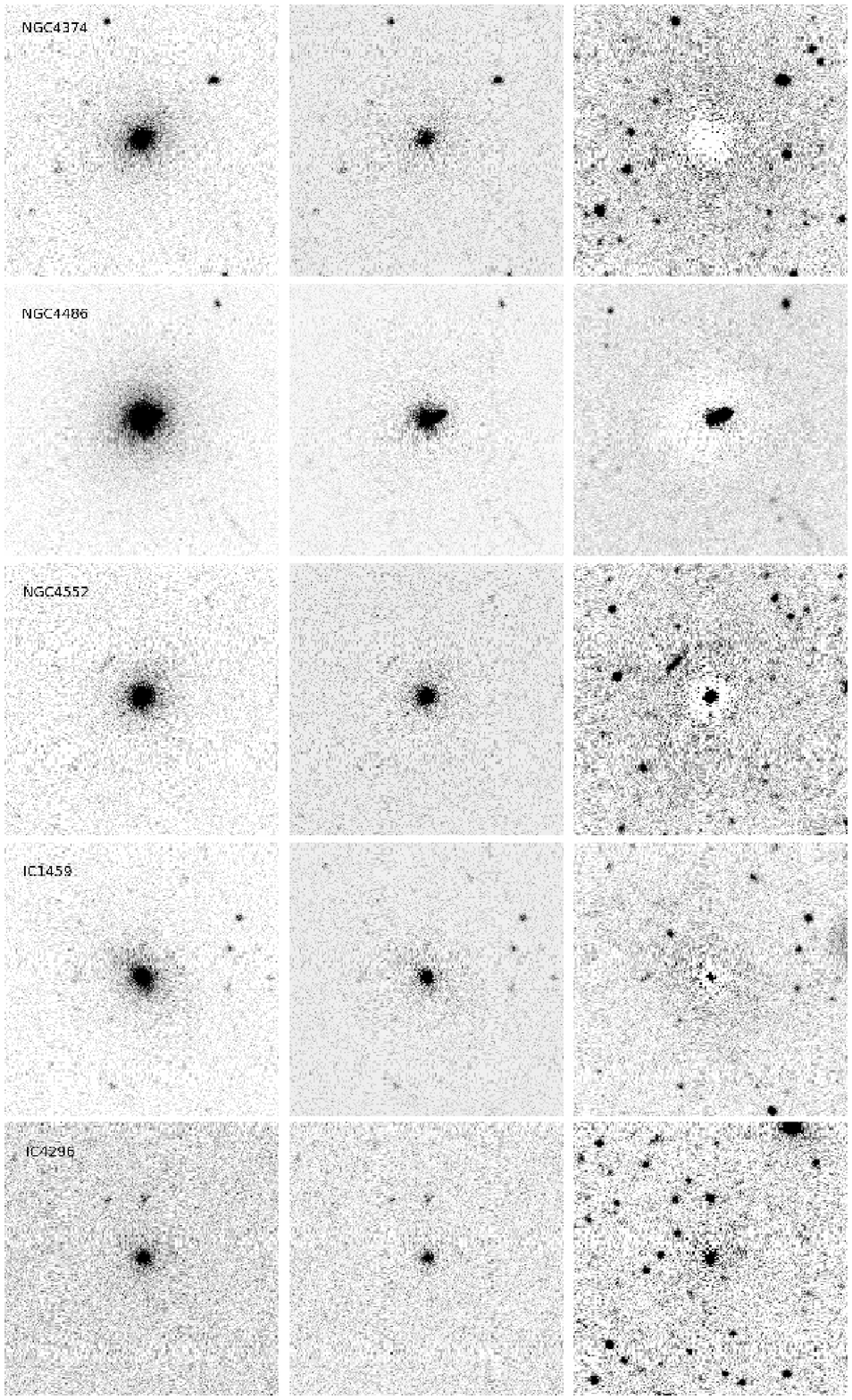} 
 \end{center} 
 \caption[]{FUV, FUV residual and NUV residual images of (from top) NGC4374, NGC4486, NGC4552, IC1459 and IC4296. Columns as in Figure B1}
 \label{fig:fig2f}
 \end{figure*}

\label{lastpage}


\begin{thebibliography}{}
\bibitem[\protect\citeauthoryear{Annibali et al.}{2006}]{2006A&A...445...79A} Annibali F., Bressan A., Rampazzo R., Zeilinger W.~W., 2006, A\&A, 445, 79 
\bibitem[Annibali et al. (2007)]{Ann07} Annibali, F., Bressan, A., Rampazzo, R., Zeilinger, W.~W. \& Danese, L. 2007,
A\&A, 463, 455 
\bibitem[Bianchi et al. (2005)]{Bia05} Bianchi, L., et al. 2005, ApJL, 619, L71
\bibitem[\protect\citeauthoryear{Brough et al.}{2007}]{2007MNRAS.378.1507B} Brough S., Proctor R., Forbes D.~A., Couch W.~J., Collins C.~A., Burke D.~J., Mann R.~G., 2007, MNRAS, 378, 1507 
\bibitem[Brown et al. (1997)]{Brown97} Brown, T.~M., Ferguson, H.~C., Davidsen, A.~F., \& Dorman, B. 1997, ApJ, 482, 685
\bibitem[Brown et al. (2000a)]{Brown00a} Brown, T.~M., Bowers, C.~W., Kimble, R.~A., \& Ferguson, H.~C. 2000a, ApJ, 529, L89
\bibitem[Brown et al. (2000b)]{Brown00b} Brown, T.~M., Bowers, C.~W., Kimble, R,~A., \& Ferguson, H.~C. 2000b, ApJ, 532, 308
\bibitem[Brown et al. (2003)]{Brown03} Brown, T.~M., Ferguson, H.~C., Smith, E., Bowers, C.~W., Kimble, R.~A., Renzini, A. \& Rich, R.~M. 2003, ApJ, 584, L69
\bibitem[Brown et al. (2008)]{Brown08} Brown, T.~M., Smith, E., Ferguson, H.~C., Sweigert, A.~V., Kimble, R.~A., \& Bowers, C. ~W. 2008, ApJ, 682, 319
\bibitem[\protect\citeauthoryear{Bruzual \& Charlot}{2003}]{2003MNRAS.344.1000B} Bruzual G., Charlot S., 2003, MNRAS, 344, 1000 
\bibitem[Burstein et al. (1988)]{Bur88} Burstein. D., Bertola, F., Buson, L.~M., Faber, S.~M., \& Lauer, T.~R. 1988, ApJ, 328, 440
\bibitem[Carter et al. (2009)]{Car09} Carter, D., et al. 2009, MNRAS, 397, 695
\bibitem[Code \& Welch (1979)]{CW79} Code, A.~D. \& Welch, G.~A. 1979, ApJ, 229, 95
\bibitem[Chuzhoy \& Loeb (2004)]{CzL04} Chuzhoy, L. \& Loeb, A. 2004, MNRAS, 349, L13
\bibitem[D'Antona \& Ventura (2007)]{DAV07} D'Antona, F. \& Ventura P. 2007, MNRAS, 379, 1431
\bibitem[Decressin et al. (2007)]{Dec07} Decressin, T., Meynet, G., Charbonnel C. Prantzos, N., \& Ekstr\"om, S. 2007,
A\&A, 464, 1029
\bibitem[De Vaucouleurs (1948)]{deV48} De Vaucouleurs, G.\ 1948, Annales d'Astrophysique, 11, 247
\bibitem[De Vaucouleurs et al. (1991)]{RC3}De Vaucouleurs G., De Vaucouleurs A., Corwin H. G., Buta R. J., Paturel G., Fouque P., 1991, 
Third Reference Catalog of Bright Galaxies, Vols. 1-3, Springer-Verlag, Berlin (RC3)
\bibitem[Di Matteo et al. (2009)]{DiM09} Di Matteo, P., Pipino, A., Lehnert, M.~D., Combes, F. \& Semelin, B. 2009,
A\&A, 499, 427
\bibitem[Donas et al. (2007)]{Don07} Donas, J., et al. 2007, ApJS, 173, 597
\bibitem[Dorman et al. (1995)]{Dor95} Dorman, B., O'Connell, R.~W. \& Rood, R.~T. 1995, ApJ, 442, 105
\bibitem[Forbes \& Thomson (1992)]{FT92} Forbes, D.A. \& Thomson, R.C. 1992, MNRAS, 254, 723
\bibitem[Gil de Paz et al. (2007)]{GdP07} Gil de Paz, A., et al. 2007, ApJS, 173, 185
\bibitem[Greggio \& Renzini (1990)]{GR90} Greggio, L. \& Renzini, A. 1990, ApJ, 364, 35
\bibitem[Han et al. (2002)]{han02} Han Z., Podsiadlowski Ph., Maxted P.~F.~L., Marsh T.~R. \& Ivanova N. 2002, MNRAS, 336, 449
\bibitem[Han et al. (2003)] {han03} Han Z., Podsiadlowski Ph., Maxted P.~F.~L. \& Marsh T.~R. 2003, MNRAS, 341, 669 
\bibitem[Han et al. (2007)]{han07} Han, Z., Podsiadlowski, Ph. \& Lynas-Gray, A.~E. 2007, MNRAS, 380, 1098
\bibitem[Han (2008)]{han08} Han, Z., 2008, A\&A, 484, L31
\bibitem[Hau et al. (1999)]{Hau99} Hau, G.~K.~T., Carter, D. \& Balcells, M. 1999, MNRAS, 306, 437
\bibitem[Jarrett et al. (2003)]{Jar03} Jarrett, T.~H., Chester, T., Cutri, R., Schneider, S.~E. \& Huchra, J.~P. 2003, AJ, 125, 525
\bibitem[Jeong et al. (2009)]{Je09} Jeong, H., et al. 2009, MNRAS, 398, 2028
\bibitem[Kaviraj et al. (2007)]{Kav07} Kaviraj, S., Sohn, S.~T., O'Connell, R.~W., Yoon, S.-J., Lee, Y.-W., \& Yi, S.~K. 2007, MNRAS, 377, 987
\bibitem[Kobayashi (2004)]{Kob04} Kobayashi, C. 2004, MNRAS, 347, 740 
\bibitem[Kobayashi \& Arimoto (1999)]{KA99} Kobayashi, C. \& Arimoto, N. 1999, ApJ, 527, 573
\bibitem[Kormendy et al. (2009)]{Kor09} Kormendy, J., Fisher, D.~B., Cornell, M.~E. \&  Bender, R. 2009, ApJS, 182, 216
\bibitem[Kuntschner et al. (2010)]{Kun10} Kuntschner, H., et al. 2010, MNRAS, 408, 97
\bibitem[La Barbera et al. (2005)]{LaB05} La Barbera, F., de Carvalho, R.~R., Gal, R.~R., Busarello, G., Merluzzi, P., Capaccioli, M. \& Djorgovski, S. G. 2005,
ApJL, 626, L19
\bibitem[La Barbera et al. (2010)]{LaB10} La Barbera, F., de Carvalho, R.~R., De La Rosa, I.~G., Gal, R.~R., Swindle, R. \& Lopes, P.~A.~A. 2010,
AJ, 140, 1528
\bibitem[Lee et al. (2007)]{Lee07} Lee, Y.-W., et al., 2005, ApJL, 621, L57
\bibitem[\protect\citeauthoryear{Lee et al.}{2009}]{2009ApJ...694..902L} Lee H.-c., et al., 2009, ApJ, 694, 902
\bibitem[Lisker \& Han (2008)]{lis08} Lisker, T. \& Han, Z. 2008, ApJ, 680, 1042
\bibitem[Loubser et al. (2009)]{Lou08} Loubser, S.~I., S\'anchez-Bl\'azquez, P.; Sansom, A.~E. \& Soechting, I.~K. 2009, MNRAS, 398, 133L
\bibitem[Loubser \& S\'anchez-Bl\'azquez (2011)]{Lou11} Loubser, S.~I. \& S\'anchez-Bl\'azquez, P. 2011, MNRAS, 410, 2679
\bibitem[Magris \& Bruzual (1993)]{MB93} Magris, G. \& Bruzual, G. 1993, ApJ, 417, 102
\bibitem[Malin (1979)]{Mal79} Malin, D.F. 1979, Nature, 277, 279
\bibitem[Malin \& Carter (1980)]{MC80} Malin, D.F. \& Carter, D. 1980, Nature 285, 643
\bibitem[Marino et al. (2011)]{Mar11} Marino, A., et al. 2010, MNRAS, 411, 311
\bibitem[Martin et al. (2005)]{Mart05} Martin, D., et al. 2005, ApJL, 619, L1
\bibitem[Norris (2004)]{Nor04} Norris, J.~E. 2004, ApJL, 612, L25
\bibitem[O'Connell et al. (1992)]{oc92} O'Connell, R.W., et al. 1992, ApJL, 395, L45
\bibitem[O'Connell (1999)]{oc99} O'Connell, R.~W. 1999, ARA\&A, 37, 603
\bibitem[Ogando et al. (2005)]{Og05} Ogando, R.~L.~C., Maia, M.~A.~G., Chiappini, C., Pellegrini, P.~S., Schiavon, R.~P. \& da Costa, L.~N. 2005,
ApJL, 632, L61
\bibitem[Ogando et al. (2008)]{Og08} Ogando, R.~L.~C., Maia, M.~A.~G., Pellegrini, P.~S. \& da Costa, L.~N. 2008, AJ, 135, 2424
\bibitem[Ogando et al. (2010)]{Og10} Ogando, R.~L.~C., Maia, M.~A.~G., Pellegrini, P.~S. \&  da Costa, L.~N. 2010, IAUS, 262, 400
\bibitem[Park \& Lee (1997)]{Park97} Park, J.-H. \& Lee, Y.-W. 1997, ApJ, 476, 28
\bibitem[Peletier et al. (1990)]{Pel90} Peletier, R.~F., Davies, R.~L., Illingworth, G.~D., Davis, L.~E. \& Cawson, M. 1990, AJ, 100, 1091
\bibitem[Peng et al. (2002)]{Peng02} Peng, C.~Y., Ho, L.~C., Impey, C.~D. \& Rix, H.-W. 2002, AJ, 124, 266
\bibitem[Peng et al. (2010)]{Peng010} Peng, C.~Y., Ho, L.~C., Impey, C.~D. \& Rix, H.-W. 2010, AJ, 139, 2097
\bibitem[Peng \& Nagai (2009)]{PN09} Peng, F. \& Nagai, D. 2009, ApJL, 705, L58
\bibitem[Percival \& Salaris (2011)]{Per11} Percival, S.~M. \& Salaris, M. 2011, MNRAS in press (DOI:10.1111/j.1365-2966.2010.18066.x)
\bibitem[Pipino \& Matteucci (2004)]{Pip04} Pipino, A. \& Matteucci, F. 2004, MNRAS, 347, 489
\bibitem[Pipino et al. (2006)]{Pip06} Pipino, A., Matteucci, F. \& Chiappini, C. 2006, ApJ, 638, 739
\bibitem[Pipino et al. (2008)]{Pip08} Pipino, A., D'Ercole, A. \& Matteucci, F. 2008, A\&A, 484, 669
\bibitem[Pipino et al. (2010)]{Pip10} Pipino, A., D'Ercole, A., Matteucci, F. \& Chiappini, C. 2010, MNRAS, 407, 1347
\bibitem[\protect\citeauthoryear{Proctor \& Sansom}{2002}]{2002MNRAS.333..517P} Proctor R.~N., Sansom A.~E., 2002, MNRAS, 333, 517 
\bibitem[\protect\citeauthoryear{Rampazzo et al.}{2005}]{2005A&A...433..497R} Rampazzo R., Annibali F., Bressan A., 
Longhetti M., Padoan F., Zeilinger W.~W., 2005, A\&A, 433, 497 
\bibitem[Rawle et al. (2008)]{Raw08} Rawle, T.~D., Smith, R.~J., Lucey, J.~R., Hudson, M.~J. \& Wegner, G.~A. 2008,
MNRAS, 385, 2097
\bibitem[\protect\citeauthoryear{Reda et al.}{2007}]{2007MNRAS.377.1772R} Reda F.~M., Proctor R.~N., Forbes D.~A., Hau G.~K.~T., Larsen S.~S., 2007, MNRAS, 377, 1772 
\bibitem[Ree et al. (2007)]{Ree07} Ree, C.~H. et al. 2007, ApJS, 173, 607
\bibitem[Reimers (1975)]{Reimers75} Reimers, D. 1975, M\'em. Soc. Roy. Sci. Li\`ege, 6th ser., 8, 369
\bibitem[\protect\citeauthoryear{S{\'a}nchez-Bl{\'a}zquez et al.}{2007}]{2007MNRAS.377..759S} S{\'a}nchez-Bl{\'a}zquez P., Forbes D.~A., Strader J., Brodie J., Proctor R., 2007, MNRAS, 377, 759 
\bibitem[\protect\citeauthoryear{Schiavon}{2007}]{2007ApJS..171..146S} Schiavon R.~P., 2007, ApJS, 171, 146
\bibitem[Schlegel et al. (1998)]{Schl88} Schlegel, D.J., Finkbeiner, D.P. \& Davis, M. 1998, ApJ,
500, 525
\bibitem[Schweizer (1980)]{Schw80} Schweizer, F. 1980, ApJ, 237, 303
\bibitem[Serra et al. (2008)]{Ser08} Serra, P., Trager, S.~C., Oosterloo, T.~A. \& Morganti, R. 2008,
A\&A, 483, 57
\bibitem[S\'ersic (1963)]{Sersic63} S\'ersic, J.~L.\ 1963, Bolet\'{\i}n de la Asociaci\'on Argentina de Astronom\'{\i}a, 6, 41
\bibitem[S\'ersic (1968)]{Sersic68} S\'ersic, J.~L.\ 1968, Atlas de Galaxias Australes (Cordoba: Observatorio Astronomico)
\bibitem[Sikkema et al. (2007)]{Sik07} Sikkema, G., Carter, D., Peletier, R.F., Balcells, M., del Burgo, C. \& Valentijn, E.A. 2007,
A\&A, 467, 1011
\bibitem[Skrutskie et al. (2006)]{Skr06} Skrutskie, M.~F. et al. 2006, AJ, 131, 1163
\bibitem[Sohn et al. (2006)]{Sohn06} Sohn, S.~T., O'Connell, R.~W., Kundu, A., Landsman, W.~B., Burstein, D., 
Bohlin, R.~C., Frogel, J.~A. \& Rose, J.~A. 2006, AJ, 131, 866
\bibitem[\protect\citeauthoryear{Spolaor et al.}{2008}]{2008MNRAS.385..675S} Spolaor M., Forbes D.~A., Proctor R.~N., Hau G.~K.~T., Brough S., 2008, MNRAS, 385, 675 
\bibitem[Spolaor et al. (2009)]{Spo09} Spolaor, M., Proctor, R.~N., Forbes, D.~A. \& Couch, W.~J. 2009, ApJL, 691, L138
\bibitem[Spolaor et al. (2010)]{Spo10} Spolaor, M., Kobayashi, C., Forbes, D.~A., Couch, W.~J. \& Hau, G.~K.~T. 2010,
MNRAS, 408, 272
\bibitem[Thomas et al. (2003)]{Tho03} Thomas, D., Maraston, C. \& Bender, R. 2003, MNRAS, 339, 897
\bibitem[Thomas et al. (2005)]{Tho05} Thomas, D., Maraston, C., Bender, R. \& Mendes de Oliveira, C. 2005,
ApJ, 621, 673
\bibitem[Tortora et al. (2010)]{Tor10} Tortora, C., Napolitano, N. R., Cardone, V. F., Capaccioli, M., Jetzer, Ph. \& Molinaro, R. 2010,
MNRAS, 407, 144 
\bibitem[Turnbull et al. (1999)]{Tur99} Turnbull, A., Bridges, T.~J. \& Carter, D. 1999, MNRAS, 307, 967
\bibitem[Wyder et al. (2005)]{Wyd05} Wyder, T.K., et al. 2005, ApJL, 619, L15
\bibitem[Yi et al. (1997)]{yi97} Yi, S.~K., Demarque, P. \& Oemler, A. Jr. 1997, ApJ, 486, 201
\bibitem[Yi et al. (1999)]{yi99} Yi, S.~K., Lee, Y.-W., Woo, J.-H., Park, J.-H., Demarque, P., \& Oemler,  A. Jr. 1999, ApJ, 513, 128
\bibitem[Yi (2008)]{yi08} Yi, S.~K. 2008, ASP Conference series, 392, 3

\end{thebibliography}
\end{document}